\documentclass[preprint, aps, prd, superscriptaddress, nofootinbib, floatfix]{revtex4-2}

\usepackage[T1]{fontenc}
\usepackage[utf8]{inputenc}
\usepackage{amsmath, amssymb, physics, bm}
\usepackage{graphicx, xcolor}
\usepackage{siunitx}
\usepackage{booktabs}
\usepackage{comment}
\usepackage{multirow}
\usepackage{caption}
\usepackage{geometry,tabularx}
\usepackage{enumitem}
\usepackage{subcaption}
\usepackage[normalem]{ulem}

\newcommand{\ws}{w_s}
\newcommand{\Omr}{\Omega_r}

\newcommand{\fend}{f_{\mathrm{end}}}
\newcommand{\fbeg}{f_{\mathrm{beg}}}

\newcommand{\Ccal}{\mathcal{C}}

\newcommand{\paperI}{\cite{BarenboimBurns:Paper1}}

\newenvironment{nota}{\par\medskip\noindent\textbf{Note.~}}{\par\medskip}

\usepackage[colorlinks=true, linkcolor=blue, citecolor=blue, urlcolor=blue]{hyperref}
\usepackage{cleveref}

\begin{document}

\flushbottom

\title{\Large Detecting Cosmological Stasis with Future Gravitational Wave Observatories}

\author{Gabriela Barenboim}
\affiliation{Instituto de F\'{i}sica Corpuscular, CSIC-Universitat de Val\`{e}ncia, Paterna 46980, Spain}
\affiliation{Departament de F\'{i}sica Te\`{o}rica, Universitat de Val\`{e}ncia, Burjassot 46100, Spain}

\author{Anne-Katherine Burns}
\affiliation{Departament de F\'{i}sica Qu\`{a}ntica i Astrof\'{i}sica (FQA), Universitat de Barcelona (UB),
  c.\ Mart\'{\i} i Franqu\`{e}s, 1, 08028 Barcelona, Spain}
\affiliation{Institut de Ci\`{e}ncies del Cosmos (ICCUB), Universitat de Barcelona (UB),
  c.\ Mart\'{\i} i Franqu\`{e}s, 1, 08028 Barcelona, Spain}

\email{gabriela.barenboim@uv.es}
\email{annekatherineburns@icc.ub.edu}

\begin{abstract}
We map the observational predictions of cosmological stasis in the inflationary
gravitational wave background (IGWB) onto the sensitivity bands of current and
planned gravitational wave detectors.  Using the closed-form piecewise spectral
template derived in the companion paper~\paperI, we generate detectability maps
for four stasis scenarios — canonical ($0\leq w_s < 1/3$), dynamical scalar ($1/3 < w_s < 1$),
vacuum-energy/matter ($-1/3 < w_s < 0$), and
vacuum-energy/radiation ($-1/3 < w_s < 1/3$) — across the frequency bands
probed by NANOGrav, SKA, LISA, DECIGO, BBO, the Einstein Telescope, and Cosmic
Explorer.  For scenarios in which the spectrum is suppressed, $w_s < 1/3$, the stasis feature is detectable by BBO in the region of $(w_s,\Delta N)$ parameter space in which $w_s\gtrsim 0.2$ for tensor-to-scalar ratios close to the Planck upper limit, r = 0.036, while to be detectable by DECIGO requires $w_s \gtrsim 0.3$. For scenarios in which the spectrum is enhanced, $w_s > 1/3$, the stasis feature is detectable by BBO across the entire
$(w_s,\Delta N)$ parameter space for tensor-to-scalar ratios of $O(0.01)$. For large values of $w_s$ close to the kination limit, the feature is detectable for r as small as $10^{-9}$. We characterize the Standard-Model (SM) $g_*$ fine
structure of the IGWB, showing that SM phase transitions introduce spectral
steps of $\approx 20\%$ (electroweak, at $\sim 2.6\times10^{-6}$~Hz) and
$\approx 53\%$ (QCD, at $\sim 3.6\times 10^{-9}$~Hz). For stasis
scenarios with end-of-stasis temperatures below the QCD scale these steps fall
inside the stasis band and constitute additional spectral features that complement
the primary signature.  Finally, we model the finite-width end-of-stasis
transition phenomenologically, demonstrating that the spectral break at $\fend$
is smoothed over a log-frequency window $\Delta N_\mathrm{trans}\times
3(1+w_s)/4$, and that the consistency relation $\Ccal^2=\Ccal^2(\alpha)$
remains testable provided $\Delta N_\mathrm{stasis}\gg \Delta N_\mathrm{trans}$,
a condition easily satisfied for all scenarios of phenomenological interest.
\end{abstract}

\maketitle

\tableofcontents

\section{Introduction}
\label{sec:intro}

Cosmological stasis is a dynamical fixed point of the early universe in which
the energy fractions $\Omega_i$ of multiple components remain exactly constant
over an extended epoch, even as the universe expands and
dilutes~\cite{Dienes:2021woi}.  Unlike the standard cosmological picture in
which radiation, matter, and vacuum energy each come to dominate in succession
with only transient periods of comparable contributions, stasis is a
non-trivial attractor in which an ongoing energy source continuously compensates
the dilution from Hubble expansion, locking the total equation of state at a
constant value $\ws = \sum_i w_i\Omega_i$.  Stasis has been realized in a
growing variety of microphysical settings: towers of decaying
species~\cite{Dienes:2021woi}, populations of primordial black holes evaporating
via Hawking radiation~\cite{Dienes:2022zgd, Dienes:2025qdw}, dynamical
scalars~\cite{Dienes:2024wnu}, thermal annihilation~\cite{Barber:2024vui,
Barber:2024izt}, field-dependent decay~\cite{Huang:2025odd}, and purely
gravitational interactions~\cite{Long:2025wjw}.  The recent argument
of~\cite{Halverson:2024oir} that stasis is a generic dynamical outcome of
multi-component cosmologies further motivates a systematic observational study.

The inflationary gravitational wave background (IGWB) provides a direct
observational window into the expansion history of the universe before BBN at energy scales
inaccessible to any other probe~\cite{Starobinsky:1979ty, Rubakov:1982df,
Boyle:2005se}.  Because each mode of the IGWB carries an imprint of the
equation of state at its horizon re-entry, a stasis epoch of duration $\Delta N$
e-folds leaves a characteristic feature in $h^2\Omega_\mathrm{GW}(f)$ across the
frequency band $[\fend, \fbeg]$ corresponding to modes that crossed the horizon
during stasis.  The attractor enforces $\ws = \mathrm{const}$, which makes the
tensor mode equation an exact Bessel equation and allows the entire spectral
distortion to be written in closed form.

In the companion paper~\paperI~we derived this closed-form template, established
the consistency relation $\Ccal^2 = \Ccal^2(\alpha)$ between the spectral slope
and the amplitude step, showed that these two quantities are independently
measurable and their combination is falsifiable without knowing $\ws$ in
advance, and validated the framework against the numerical PBH-stasis
calculation of~\cite{Dienes:2022zgd}.  The present paper takes the observational
step: we ask which current and planned detectors can actually detect the stasis
signature, what the minimum tensor-to-scalar ratio $r$ is for a stasis epoch to be detectable in each detector, and
what additional spectral features distinguish the stasis scenario from competing
models.

The main results of this paper are as follows.
\begin{itemize}[leftmargin=1.5em]
  \item \textbf{Detectability of Suppression Scenarios:}
    For scenarios in which the spectrum is suppressed, $w_s < 1/3$, BBO can detect the stasis feature for values of $w_s \gtrsim 0.2$ for values of r close to the Planck upper limit, 0.036; while DECIGO requires $w_s \gtrsim 0.3$; LISA, ET, and CE are competitive only for $r$ well above current upper limits.
    The primary observational target for stasis detection when $w_s < 1/3$ in the IGWB is therefore BBO, in the decade $0.01$--$1\,\mathrm{Hz}$. 

    \item \textbf{Detectability of Enhancement Scenarios:}
    For scenarios in which the spectrum is enhanced, $w_s > 1/3$, both BBO and DECIGO can detect the stasis feature across the entire $(w_s, \Delta N)$ parameter space for values of r of $O(0.01)$. For values of $w_s$ close to the kination limit, BBO can detect the stasis features for r values as small as $\sim10^{-9}$. For stasis configurations with the longest durations thereby spanning the widest frequency bands, LISA and CE are competitive when r is of $O(0.01)$ for $w_s \gtrsim 0.55$ and ET is competitive when r is of $O(0.01)$ for $w_s \gtrsim 0.65$.

  \item \textbf{$g_*$ fine structure:}
    The Standard-Model phase transitions at the electroweak ($T_\mathrm{EW}
    \sim 100\,\mathrm{GeV}$) and QCD ($T_\mathrm{QCD}\sim 150\,\mathrm{MeV}$)
    scales suppress the spectrum at frequencies above each transition by approximately $20\%$ (electroweak) and
    $53\%$ (QCD) relative to the spectrum below at the
    corresponding frequencies $f_\mathrm{EW}\approx 2.6\times 10^{-6}\,\mathrm{Hz}$
    and $f_\mathrm{QCD}\approx 3.6\times 10^{-9}\,\mathrm{Hz}$.  For stasis
    scenarios with $T_\mathrm{end}\lesssim T_\mathrm{QCD}$, both steps fall inside
    the stasis band and are superimposed on the power-law tilt.

  \item \textbf{Smooth transition:}
    The finite-width transition out of stasis smooths the spectral break at $\fend$
    over a log-frequency window $\Delta \ln f = 3(1+w_s)/4\cdot\Delta N_\mathrm{trans}$,
    leaving a measurable shoulder whose width encodes the transition microphysics.
    The consistency relation remains intact provided $\Delta N_\mathrm{stasis}\gg
    \Delta N_\mathrm{trans}$.
\end{itemize}

This paper is organized as follows.
Section~\ref{sec:recap} summarizes the key results from~\paperI~that are used
in the rest of the paper.
Section~\ref{sec:gstar} derives the $g_*$ fine structure of the IGWB and its
interplay with the stasis template.
Section~\ref{sec:detectability} presents the detectability maps for the two
stasis scenarios across three representative frequency bands, with detectability maps for two additional scenarios appearing in the appendix.
Section~\ref{sec:transition} develops the phenomenological treatment of the
smooth end-of-stasis transition.
We conclude in Sec.~\ref{sec:discussion}.

\section{Summary of the stasis GW template}
\label{sec:recap}

We summarize the results of~\paperI~that enter the detectability analysis.
The full derivations are given there; here we collect only the operational
formulae.

\subsection{The piecewise spectral template}
\label{sec:template_recap}

Because the stasis attractor enforces a constant equation of state $\ws$, the
expansion during stasis is an exact power law in conformal time and the tensor
mode equation reduces to a Bessel equation with index
\begin{equation}
  \nu(\ws) = \frac{3(1-\ws)}{2(1+3\ws)}.
  \label{eq:nu}
\end{equation}
The sub-horizon WKB amplitude acquires the Bessel coefficient
\begin{equation}
  \Ccal^2(\nu)
  = \left[\frac{2^{\nu+1/2}\,\Gamma(\nu+1)}{\sqrt{\pi}\,\beta^\beta}\right]^2,
  \qquad
  \beta \equiv \frac{2}{1+3\ws},
  \label{eq:C2}
\end{equation}
which satisfies $\Ccal^2(1/2) = 1$ (radiation domination) and
$\Ccal^2(3/2) = 9/16$ (matter domination).  The spectral distortion relative to
the radiation-dominated baseline is
\begin{equation}
  \alpha(\ws) = \frac{2(3\ws-1)}{1+3\ws},
  \label{eq:alpha}
\end{equation}
which is negative, corresponding to a suppression or ``notch'' for canonical stasis $\ws < 1/3$
and positive, corresponding to an enhancement for $\ws > 1/3$.

The IGWB spectrum is piecewise:
\begin{equation}
  h^2\Omega_\mathrm{GW}(f)
  = h^2\Omega_\mathrm{GW}^{(\mathrm{RD})}(f)\times
  \begin{cases}
    1
    & f < \fend,\\[6pt]
    \Ccal^2(\nu)\,\bigl(f/\fend\bigr)^{\alpha(\ws)}
    & \fend < f < \fbeg,\\[6pt]
    \bigl(\fbeg/\fend\bigr)^{\alpha(\ws)}
    & f > \fbeg,
  \end{cases}
  \label{eq:template}
\end{equation}
where $h^2\Omega_\mathrm{GW}^{(\mathrm{RD})}$ is the standard flat IGWB
(see Sec.~\ref{sec:normalization} below), and the lower break frequency is
\begin{equation}
  \fend = \frac{T_0\,T_\mathrm{end}}{2\sqrt{90}\,\bar{M}_\mathrm{Pl}}
    \sqrt{g_*(T_\mathrm{end})}
    \left(\frac{g_{*s,0}}{g_{*s}(T_\mathrm{end})}\right)^{\!1/3}
    \frac{1}{\hbar}
  \;\approx\; 2.65\times 10^{-8}\,\mathrm{Hz}
    \left(\frac{T_\mathrm{end}}{1\,\mathrm{GeV}}\right)
    \left(\frac{g_*(T_\mathrm{end})}{106.75}\right)^{\!1/6},
  \label{eq:fend}
\end{equation}

where $T_0$ is the CMB temperature today, $T_\mathrm{end}$ is the final temperature of the stasis epoch, $\bar{M}_\mathrm{Pl}$ is the reduced Planck mass, and the approximate form assumes $g_* \approx g_{*s}$.  The upper break
frequency is $\fbeg = \fend\,\exp\!\bigl[(1+3\ws)\Delta N/2\bigr]$, and the
stasis duration is $\Delta N = \ln(\fbeg/\fend)\cdot 2/(1+3\ws)$.  Table~\ref{tab:C_val}
collects $\Ccal^2$ for several representative equations of state.

\renewcommand{\arraystretch}{1.25}
\begin{table}[h]
\centering
\setlength{\tabcolsep}{12pt}
\begin{tabular}{lcccc}
\toprule
Era & $\ws$ & $\alpha$ & $\nu$ & $\Ccal^2$ \\
\midrule
Matter domination (MD)   & $0$    & $-2$    & $3/2$  & $0.5625$ \\
Stasis ($\ws = 1/6$)     & $1/6$  & $-2/3$  & $5/6$  & $0.8304$ \\
Radiation domination (RD)& $1/3$  & $0$     & $1/2$  & $1.0000$ \\
Kination                 & $1$    & $+1$    & $0$    & $4/\pi$  \\
\bottomrule
\end{tabular}
\caption{Spectral tilt $\alpha$, Bessel index $\nu$, and amplitude coefficient
  $\Ccal^2(\nu)$ for several reference equations of state.
  The value $\Ccal^2 = 1$ is recovered for radiation domination,
  and $\Ccal^2 < 1$ for all $\ws < 1/3$ (canonical stasis, matter domination).}
\label{tab:C_val}
\end{table}

\subsection{IGWB normalization}
\label{sec:normalization}

For a nearly scale-invariant primordial tensor spectrum with tensor-to-scalar
ratio $r$ evaluated at the pivot scale $k_* = 0.05\,\mathrm{Mpc}^{-1}$, the
flat IGWB amplitude in the radiation-dominated limit is
\begin{equation}
  h^2\Omega_\mathrm{GW}^{(\mathrm{RD})} \simeq
    \frac{3}{128}\,h^2\Omega_{r,0}\,r\,A_s
    \left(\frac{g_*}{g_{*0}}\right)^{\!-1/3}
  \approx 6.8\times 10^{-18}\left(\frac{r}{0.01}\right)
    \left(\frac{g_*}{106.75}\right)^{\!-1/3},
  \label{eq:OmRD}
\end{equation}
where $h^2\Omega_{r,0}\approx 4.17\times 10^{-5}$, $A_s\approx
2.1\times 10^{-9}$~\cite{Planck:2018jri}, and $g_{*0} \equiv g_*(T_\mathrm{beg})$
is evaluated at the horizon-entry temperature; the approximate numerical form
normalizes to the SM value $g_* = 106.75$ (valid for $T_\mathrm{beg}\gg T_\mathrm{EW}$).  The current upper limit
$r < 0.036$ (Planck + BICEP/Keck~\cite{BICEPKeck:2021gln}) sets
$h^2\Omega_\mathrm{GW}^{(\mathrm{RD})}\lesssim 2.4\times 10^{-17}$, well
below the sensitivities of LISA and ET but within reach of BBO and DECIGO
(see Sec.~\ref{sec:detectability}).

\subsection{The consistency relation}
\label{sec:consistency_recap}

As derived in~\paperI, the spectral tilt $\alpha$ and the amplitude step
$\Ccal^2$ are both determined by a single underlying quantity $\ws$, and their
relationship traces a one-dimensional curve
$\Ccal^2 = \Ccal^2(\alpha)$ in the $(\alpha,\Ccal^2)$ plane, as shown in the following equation.

\begin{equation}
\Ccal^2 = \Ccal^2\!\left(\alpha\right) \equiv \left[\frac{2^{\nu(\alpha)+1/2}\,\Gamma(\nu(\alpha)+1)}{\sqrt{\pi}\,\beta(\alpha)^{\beta(\alpha)}}\right]^2, \quad \nu = \frac{1-\alpha}{2}, \quad \beta = \frac{2-\alpha}{2}.
\label{eq:consistency_curve}
\end{equation}

This curve is a universal Bessel-function relation satisfied by every
constant-$w$ era, regardless of the underlying microphysical mechanism.
Because $\alpha$ can be measured independently from the slope of the stasis
band and $\Ccal^2$ can be measured from the amplitude step at $\fend$, the
curve provides a falsifiable internal consistency test:
a data point that lands off the curve indicates either a time-varying
equation of state, a multi-component fluid, or systematic uncertainties in the
detector model.

\section{The \texorpdfstring{$g_*$}{g*} fine structure}
\label{sec:gstar}

The IGWB is not perfectly flat even in the absence of stasis.  As modes of
different frequency crossed the cosmological horizon at different temperatures,
they experienced different effective numbers of relativistic degrees of freedom
$g_*(T)$.  The resulting $g_*$-induced spectral modulation is a known
prediction of Standard Model thermodynamics and constitutes a background
``fine structure'' on top of which the stasis imprint sits.  In this section
we characterize this fine structure, identify which frequency bands are affected,
and discuss its interplay with the stasis template.

\subsection{Standard-Model phase transitions and spectral steps}
\label{sec:gstar_steps}

The energy density of the IGWB for modes entering the horizon at temperature
$T$ scales as~\cite{Watanabe:2006qe, Boyle:2005se}
\begin{equation}
  h^2\Omega_\mathrm{GW}^{(\mathrm{RD})}(f)
  \propto
  g_*(T(f))^{-1/3},
  \label{eq:gstar_scaling}
\end{equation}
where $T(f)$ is the temperature at horizon entry for mode $f$.  The
proportionality constant is absorbed into the normalization of
eq.~\eqref{eq:OmRD}, which is evaluated at $g_* = 106.75$.  As $g_*(T)$
decreases through a phase transition, modes that entered the horizon just
below the transition temperature see a smaller $g_*$ and therefore have a
\emph{larger} $h^2\Omega_\mathrm{GW}$: the spectrum steps up when moving from
high to low frequency through a transition.  The step amplitude is set by the
ratio of $g_*$ values on each side and is independent of normalization
convention.

The Standard Model predicts two relevant transitions in the cosmological
frequency range:
\begin{enumerate}[leftmargin=1.5em]
  \item \textbf{Electroweak crossover} ($T_\mathrm{EW}\approx 100\,\mathrm{GeV}$):
    $g_*$ drops from $106.75$ (above) to $\approx 61.75$ (below) over time, producing
    a spectral step of
    \begin{equation}
      \frac{h^2\Omega_\mathrm{GW}(f < f_\mathrm{EW})}
           {h^2\Omega_\mathrm{GW}(f > f_\mathrm{EW})}
      = \left(\frac{106.75}{61.75}\right)^{\!1/3} \approx 1.20,
      \label{eq:EW_step}
    \end{equation}
    at the frequency
    \begin{equation}
      f_\mathrm{EW}
      \approx 2.65\times 10^{-8}\,\mathrm{Hz}\times(T_\mathrm{EW}/\mathrm{GeV})
      \approx 2.6\times 10^{-6}\,\mathrm{Hz}.
      \label{eq:fEW}
    \end{equation}

  \item \textbf{QCD crossover} ($T_\mathrm{QCD}\approx 150\,\mathrm{MeV}$):
    $g_*$ drops from $\approx 61.75$ (above) to $\approx 17.25$ (below,
    after quark--hadron transition and pion decoupling), producing a step of
    \begin{equation}
      \frac{h^2\Omega_\mathrm{GW}(f < f_\mathrm{QCD})}
           {h^2\Omega_\mathrm{GW}(f > f_\mathrm{QCD})}
      = \left(\frac{61.75}{17.25}\right)^{\!1/3} \approx 1.53,
      \label{eq:QCD_step}
    \end{equation}
    at
    \begin{equation}
      f_\mathrm{QCD}
      \approx 2.65\times 10^{-8} ~\mathrm{Hz} \times 0.150 \times (g_*/106.75)^{1/6}\,
      \approx 3.6\times 10^{-9}\,\mathrm{Hz}
      \label{eq:fQCD}
    \end{equation}
\end{enumerate}
for $g_* = 61.75$. These steps are fixed, known features of the IGWB that do not depend on the
stasis parameters.  In the context of a stasis search, they can serve as
calibration points or, when they fall inside the stasis band, as complicating
features that must be modeled.

The two transitions above are not equally sharp.  The QCD crossover is a genuinely rapid transition on cosmological time scales: the quark-hadron
transition and pion decoupling occur over a narrow temperature window,
$\Delta T/T \sim \mathcal{O}(1)$, so modeling it as an instantaneous step
in $g_*(T)$, as in eq.~\eqref{eq:QCD_step}, is an accurate
approximation~\cite{Laine:2015kra}.  The electroweak transition is
different in character.  As the plasma cools through $T\sim 100$--$200\,
\mathrm{GeV}$, the top quark, Higgs boson, $W^\pm$, $Z$, and bottom quark
become non-relativistic in succession rather than simultaneously, with
their masses spanning $m_b\approx 4.2\,\mathrm{GeV}$ to
$m_t\approx 173\,\mathrm{GeV}$, roughly two decades in temperature.  The
corresponding change in $g_*(T)$ near $T_\mathrm{EW}$ is therefore not a
single sharp step but a smooth roll-off spread over the corresponding range
in horizon-entry frequency, $f\sim 10^{-7}$--$10^{-5}\,\mathrm{Hz}$ by
eq.~\eqref{eq:fend}.  The instantaneous-step treatment of
eq.~\eqref{eq:EW_step} should accordingly be understood as a schematic
approximation to this smoother feature: it correctly captures the total
amplitude change and the characteristic frequency scale, but not the
detailed shape of the transition. In the figures we replace the step in eq.~\eqref{eq:EW_step} with the smoothed profile obtained by
integrating eq.~\eqref{eq:gstar_scaling} against the full SM $g_*(T)$
table~\cite{Laine:2015kra} while the QCD step remains sharp.  This
refinement does not affect the order-of-magnitude conclusions of this
section, but it does change the detailed shape of the high-frequency feature,
which is now visible as a broad shoulder rather than a step in all figures in which the EW frequency range is visible.

\subsection{Correction to $\fend$ near the SM transitions}
\label{sec:gstar_fend}

The break-frequency formula of eq.~\eqref{eq:fend} includes a factor
$g_*(T_\mathrm{end})^{1/6}$ that depends on the SM degrees of freedom at the
end of stasis.  For $T_\mathrm{end}$ well above the EW scale, $g_* = 106.75$
and the approximate formula is exact.  Near or below the SM transitions, however,
the correct value of $g_*(T_\mathrm{end})$ deviates significantly:

\begin{itemize}[leftmargin=1.5em]
  \item $T_\mathrm{end}\approx 150\,\mathrm{MeV}$ (QCD scale):
    $g_* \approx 61.75$, giving an $8.7\%$ downward correction to $\fend$
    relative to the naive $g_* = 106.75$ estimate.
  \item $T_\mathrm{end}\approx 55\,\mathrm{MeV}$:
    $g_* \approx 10.75$, giving a $32\%$ downward correction.
    The naive $g_* = 106.75$ formula overestimates $\fend$ by nearly $50\%$.
\end{itemize}

When performing a stasis template fit to data, the correct
$g_*(T_\mathrm{end})$ must therefore be inserted into eq.~\eqref{eq:fend},
using tabulated SM $g_*(T)$ values~\cite{Laine:2015kra}.  Equivalently,
$T_\mathrm{end}$ can be treated as a free parameter whose relationship to
$\fend$ is computed from the full SM thermodynamic history.

\subsection{Spectral steps within the stasis band}
\label{sec:gstar_instasis}

The most striking effect of $g_*$ fine structure occurs when a SM transition
temperature $T_\mathrm{trans}$ falls within the range
$[T_\mathrm{end}, T_\mathrm{beg}]$ probed by the stasis band.  In that case, the
transition frequency $f_\mathrm{trans}$ lies inside $[\fend, \fbeg]$, and the
otherwise smooth power-law tilt $\alpha$ in the stasis band is interrupted by a $g_*$-induced step. In Section \ref{sec:setup}, we discuss this impact in detail for each of the configurations we consider here. These steps can be seen in Fig.~\ref{fig:ws_range2} (see §\ref{sec:setup}).

\subsection{Practical impact on parameter estimation}
\label{sec:gstar_practical}

The $g_*$ fine structure modifies the parameter-estimation problem for frequency bands where internal $g_*$ steps are present in which stasis occurs in two important ways.

First, the internal steps at $f_\mathrm{QCD}$ and $f_\mathrm{EW}$ break
the simple power law in the stasis band.  A naive fit of $\alpha$ from the
slope across the full stasis band would be biased by the presence of the
steps; instead, the slope should be fit over segments between the steps
($\fend < f < f_\mathrm{QCD}$, $f_\mathrm{QCD} < f < f_\mathrm{EW}$,
and $f_\mathrm{EW} < f < \fbeg$), each of which still has slope $\alpha(\ws)$,
and the fit should explicitly include the step amplitudes as predicted
by the SM thermal history.

Second, the two steps which occur at fixed, theoretically known frequencies serve as
internal calibration points.  The same $g_*$ history that produces the steps
also sets the overall IGWB normalization via eq.~\eqref{eq:gstar_scaling},
providing a self-consistency check between the step amplitudes and the
continuum shape.  A mismatch between the observed step amplitude and the
SM prediction would signal new physics, e.g., additional light degrees of
freedom active near the relevant transition temperature.

As discussed in Section \ref{sec:setup}, for the frequency bands in which stasis occurs where no internal $g_*$ steps are present, the stasis
template in its idealized form~\eqref{eq:template} is directly applicable
with only the overall normalization correction from eq.~\eqref{eq:gstar_scaling}.
The $g_*$ correction to $\fend$ (Sec.~\ref{sec:gstar_fend}) should nonetheless
be applied when inverting an observed $\fend$ to a physical temperature
$T_\mathrm{end}$.

\begin{nota}
  The $g_*$ fine structure of the IGWB is a predicted feature of Standard Model
  thermodynamics.  In the context of a stasis search, the steps at
  $f_\mathrm{QCD}$ and $f_\mathrm{EW}$ are \emph{not} free parameters: their
  frequencies and amplitudes are fixed by the SM.  A detection of stasis in a
  frequency band in which internal $g_*$ steps are present should therefore simultaneously reproduce the known $g_*$ steps at
  the correct frequencies, providing a non-trivial cross-check of both the
  stasis hypothesis and Standard Model thermodynamics.
\end{nota}

\section{Detectability}
\label{sec:detectability}

\subsection{Setup}
\label{sec:setup}

We overlay the stasis GW template of eq.~\eqref{eq:template} on the
power-law integrated sensitivity (PLS) curves~\cite{Schmitz:2020rag}
of the following instruments:
NANOGrav~\cite{NANOGrav:2020bcs},
EPTA~\cite{EPTA:2021crs},
SKA~\cite{Janssen:2014dea},
LISA~\cite{LISA:2017pwj},
DECIGO~\cite{Kawamura:2011zz},
BBO~\cite{Corbin:2005ny},
Einstein Telescope (ET-D)~\cite{Punturo_2010},
and Cosmic Explorer (CE)~\cite{Reitze:2019iox}.
We also show the NANOGrav 15-year posterior~\cite{NANOGrav:2023gor}
as a reference blue band.

All spectra are shown relative to the IGWB reference baseline at $r = 0.01$
(grey dashed line), which gives $h^2\Omega_\mathrm{GW}^{(\mathrm{RD})}\approx
6.8\times 10^{-18}$ at high frequency.  For each scenario we choose three representative
frequency-band configurations:
\begin{itemize}[leftmargin=1.5em]
  \item \textbf{Band~I:} $\fend = 10^{-4}\,\mathrm{Hz}$,
    $\fbeg = 10^{-1}\,\mathrm{Hz}$, squarely in the LISA band.
    Corresponds to $T_\mathrm{end}\approx 3.8\,\mathrm{TeV}$,
    well above the electroweak scale. Here, the stasis band has $g_* = 106.75$
    throughout.  No internal $g_*$ steps; only the overall normalization
    correction~\eqref{eq:gstar_scaling} applies.
  \item \textbf{Band~II:} $\fend = 10^{-9}\,\mathrm{Hz}$,
    $\fbeg = 10^{-2}\,\mathrm{Hz}$, spanning the PTA-to-LISA range.
    Corresponds to $T_\mathrm{end}\approx 55\,\mathrm{MeV}$,
    below the QCD crossover. Here, $f_\mathrm{QCD}\approx 3.6\times 10^{-9}\,\mathrm{Hz}$ and
    $f_\mathrm{EW}\approx 2.6\times 10^{-6}\,\mathrm{Hz}$ both fall inside
    $[\fend, \fbeg] = [10^{-9}, 10^{-2}]\,\mathrm{Hz}$.
    The stasis band therefore contains \emph{two additional spectral steps}
    of $\approx 53\%$ (QCD) and $\approx 20\%$ (EW) superimposed on the
    power-law tilt $\alpha(\ws)$.  These steps are at known, fixed
    frequencies and are distinct from the stasis breaks at $\fend$ and $\fbeg$.
    Note, however, that $f_\mathrm{QCD} \approx 3.6\,\fend$ for this
    configuration, so the QCD step sits close to the lower break; a joint
    template fit including both the stasis break at $\fend$ and the
    $g_*$-induced step at $f_\mathrm{QCD}$ is necessary to avoid bias.
  \item \textbf{Band~III:} $\fend = 10^{-2}\,\mathrm{Hz}$,
    $\fbeg = 10^{3}\,\mathrm{Hz}$, spanning DECIGO/BBO to ET/CE.
    Corresponds to $T_\mathrm{end}\approx 380\,\mathrm{TeV}$. As in Band~I, both transition frequencies lie below $\fend$.  No internal
    $g_*$ steps.
\end{itemize}

These three bands have been intentionally chosen to probe distinct thermal scales rather than to span equal frequency ranges. Table~\ref{tab:bands} summarizes these configurations and the corresponding
thermal scales.

\begin{table}[h]
\centering
\renewcommand{\arraystretch}{1.25}
\setlength{\tabcolsep}{12pt} 
\begin{tabular}{lcccc}
\toprule
Band & $\fend$ & $\fbeg$ & $T_\mathrm{end}$ & Primary detector(s) \\
\midrule
I   & $10^{-4}$\,Hz & $10^{-1}$\,Hz & $\sim 3.8$\,TeV    & LISA \\
II  & $10^{-9}$\,Hz & $10^{-2}$\,Hz & $\sim 55$\,MeV     & NANOGrav, SKA, LISA \\
III & $10^{-2}$\,Hz & $10^{3}$\,Hz  & $\sim 380$\,TeV    & BBO, DECIGO, ET, CE \\
\bottomrule
\end{tabular}
\caption{Frequency-band configurations used in the detectability plots.
  The break-frequency-to-temperature mapping uses
  eq.~\eqref{eq:fend} with $g_*$ fine structure corrections included.
  The $g_*$ correction for Band~II is discussed in Sec.~\ref{sec:gstar}.}
\label{tab:bands}
\end{table}

\begin{figure}[t]
  \centering
  \includegraphics[width=1\linewidth]{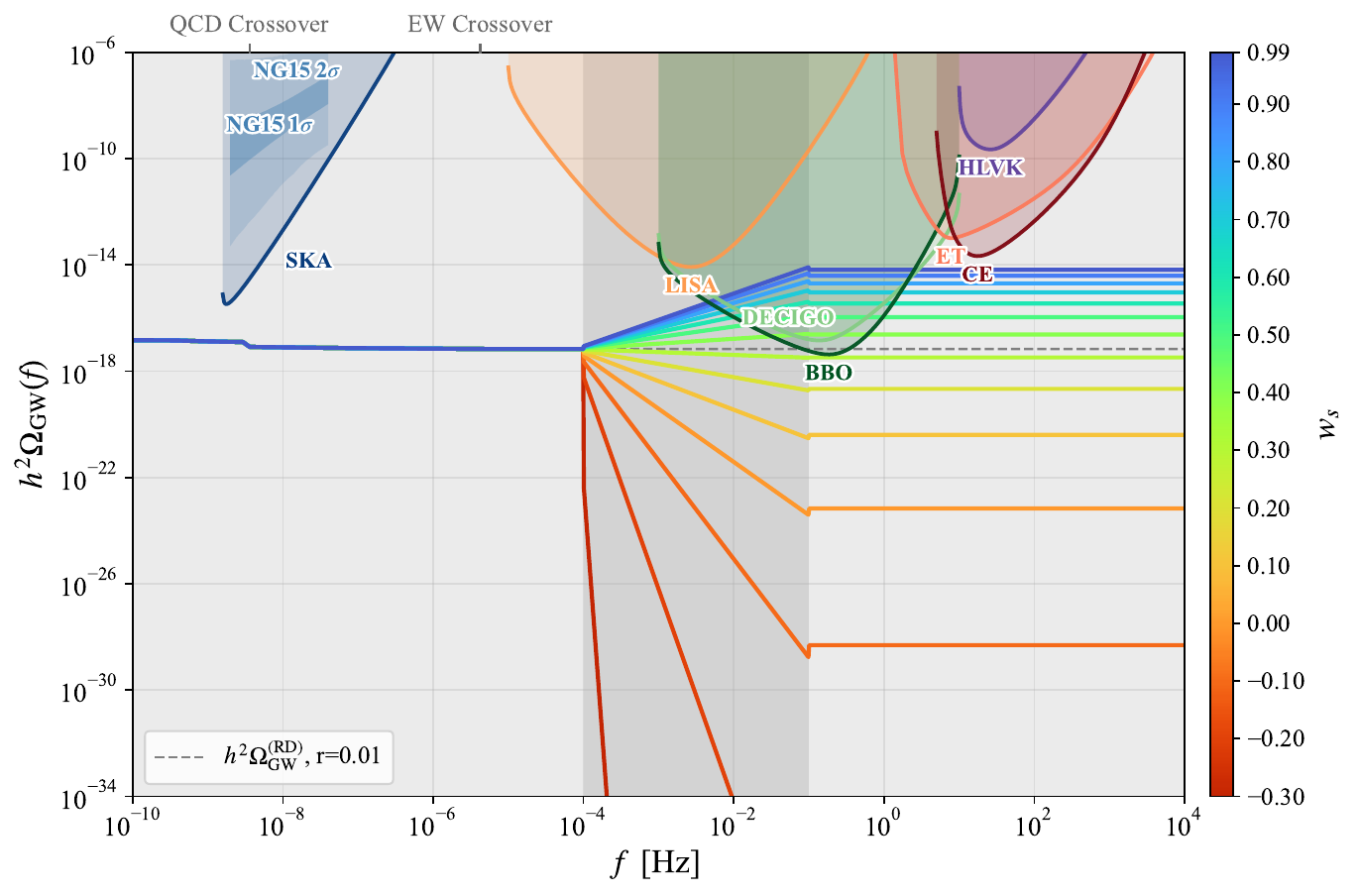}
  \caption{Stasis-modified IGWB for several equations of state
    $\ws\in[-1/3,1]$ in Band~I
    ($\fend=10^{-4}\,\mathrm{Hz}$, $\fbeg=10^{-1}\,\mathrm{Hz}$).
    Shaded regions: PLS curves~\protect\cite{Schmitz:2020rag} with $T_{obs}$ = 4 years and $\rho_{thr}$ = 1.
    Grey dashed: RD baseline at $r = 0.01$. The \texorpdfstring{$g_*$}{g*} fine structure, discussed in Section \ref{sec:gstar} is incorporated in all spectra.}
  \label{fig:ws_range}
\end{figure}

\begin{figure}[t]
  \centering
  \includegraphics[width=1\linewidth]{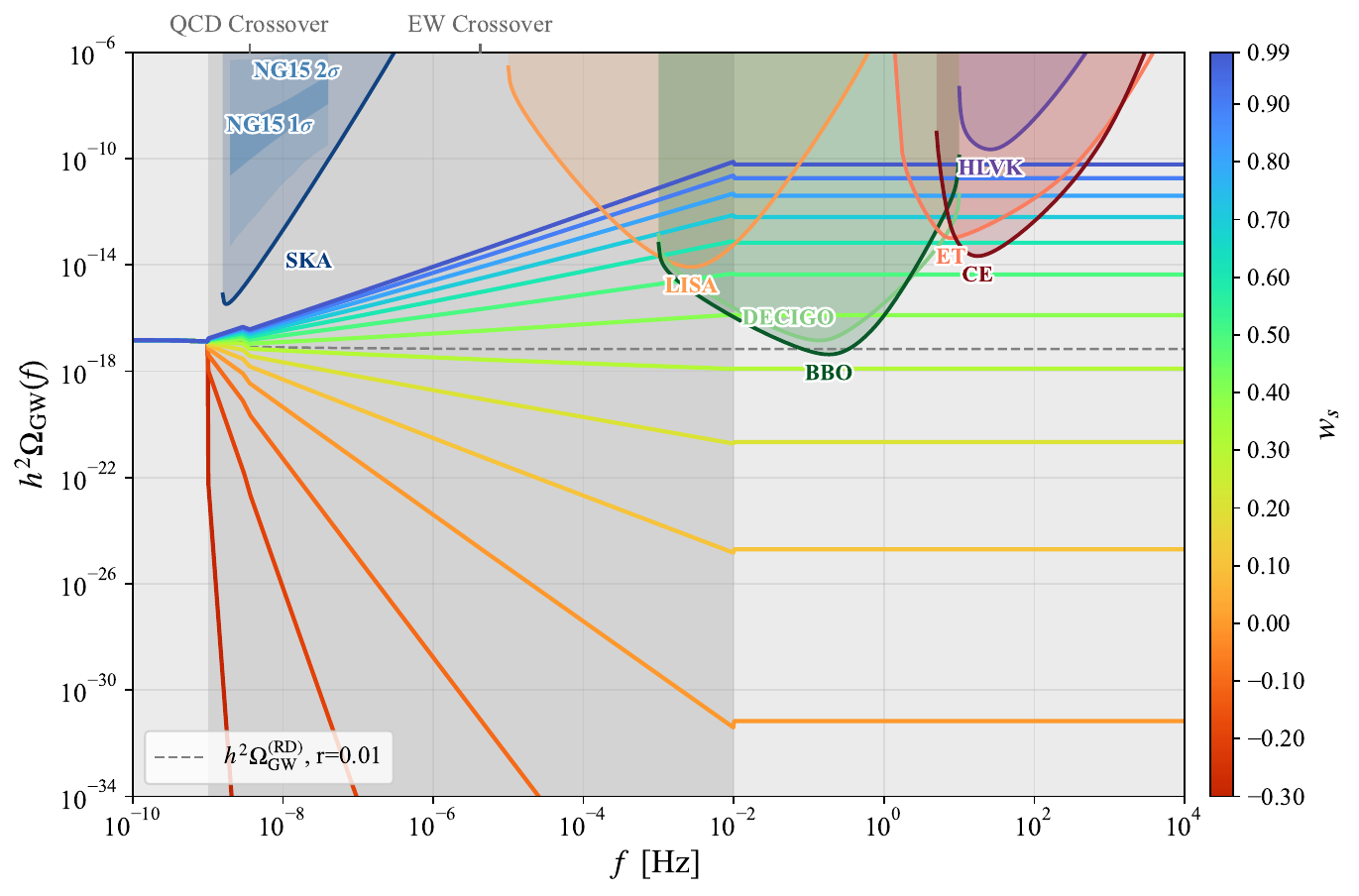}
  \caption{Same as Fig.~\ref{fig:ws_range} for Band~II
    ($\fend=10^{-9}\,\mathrm{Hz}$, $\fbeg=10^{-2}\,\mathrm{Hz}$).
    The $g_*$ fine structure of Sec.~\ref{sec:gstar} adds spectral steps
    at $f_\mathrm{QCD}\approx 3.6\times 10^{-9}\,\mathrm{Hz}$ and
    $f_\mathrm{EW}\approx 2.6\times 10^{-6}\,\mathrm{Hz}$, which fall
    inside the stasis band for this configuration.}
  \label{fig:ws_range2}
\end{figure}

\begin{figure}[t]
  \centering
  \includegraphics[width=1\linewidth]{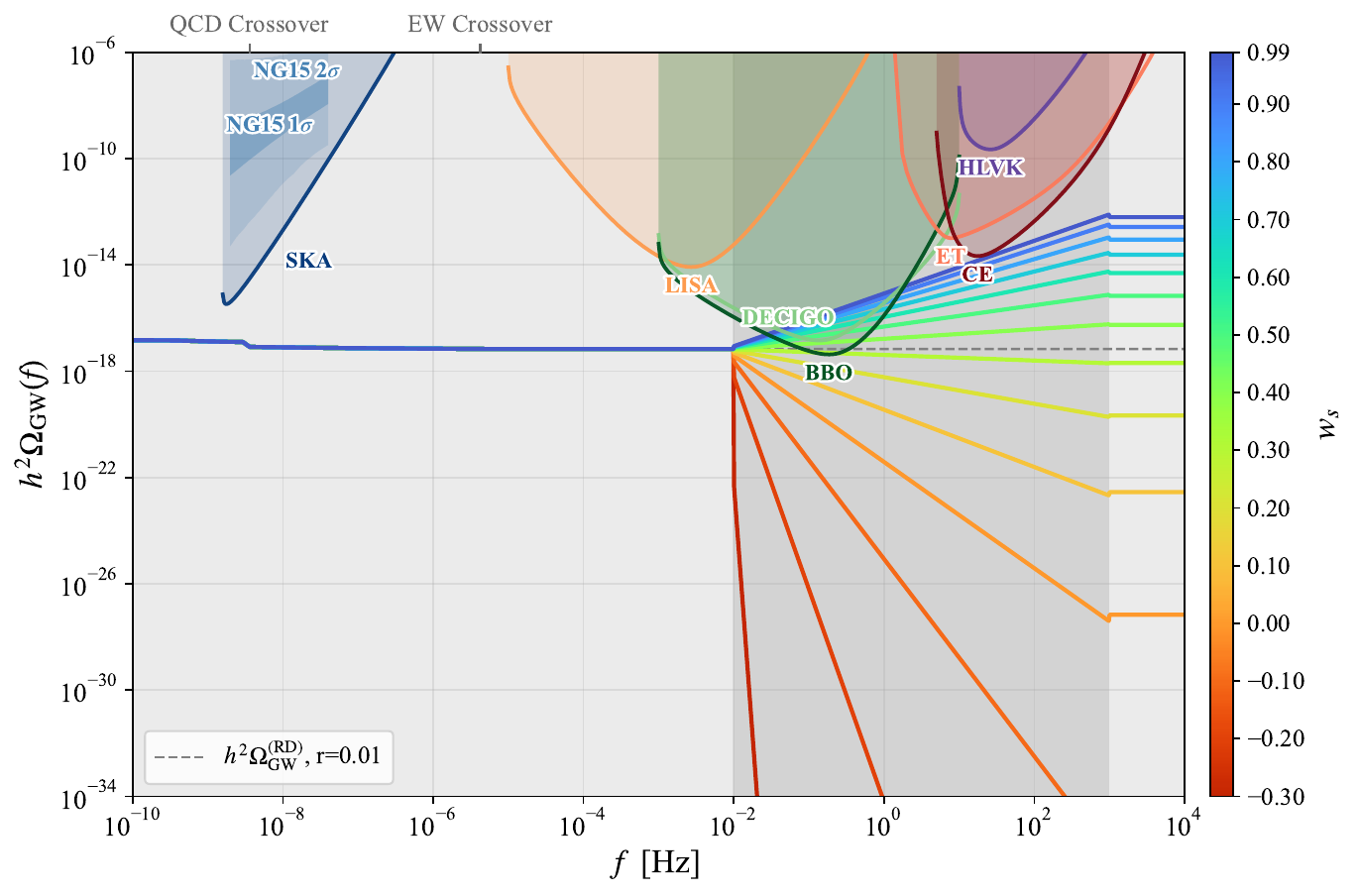}
  \caption{Same as Fig.~\ref{fig:ws_range} for Band~III
    ($\fend=10^{-2}\,\mathrm{Hz}$, $\fbeg=10^{3}\,\mathrm{Hz}$).
    This configuration places the stasis feature in the BBO/DECIGO-to-ET
    frequency range, which has the best projected sensitivity to the IGWB
    among the considered detectors.}
  \label{fig:ws_range3}
\end{figure}

Figures~\ref{fig:ws_range},~\ref{fig:ws_range2}, and~\ref{fig:ws_range3} show the spectral shape for a range of
$\ws$ values across all three bands, illustrating how the spectral tilt and
amplitude step evolve across the parameter space. It is important to note that for some stasis epochs, the frequency band during which stasis occurs might be such that an enhancement or suppression, and therefore the spectral distortion, $\alpha$, is measurable by a detector but the amplitude step, $\Ccal^2$ is not. While a measured slope, $\alpha$ is a hint that a stasis epoch occurred, the consistency relation $\Ccal^2 = \Ccal^2(\alpha)$ of Eq. \ref{eq:consistency_curve} is what makes the test specific to stasis,
and that requires both quantities to be independently measurable. In the Band~III case as shown in Fig. \ref{fig:ws_range3}, for $w_s > 0.3$, the suppression or enhancement is measurable, but the amplitude step is not meaning that if stasis were to have occurred during this time, our projected future detector landscape would be unable to verify it through the consistency relation. 

\subsection{Detectability maps}
\label{sec:maps}

In Sec.~\ref{sec:setup} we fix both the tensor-to-scalar ratio, $r$, and the stasis duration, $\Delta N$ and display the resulting shape. Here, we show for each point on the $(\ws, r)$ plane the length of time a stasis epoch must last for its imprint to be resolvable at all. To do this, we require that the flat region below $\fend$, which fixes $\Ccal^2$, or the elevated region above $\fbeg$, which fixes $\Ccal^2(\fbeg/\fend)^\alpha$ along with some part of the tilted stasis band must  cross the PLS curve of the same instrument. Requiring only that the tilted stasis band $[\fend,\fbeg]$ rise above the PLS curve is sufficient to measure the slope $\alpha$ and hence $\ws$, but as emphasized in Sec.~\ref{sec:setup} a measured slope alone is a hint rather than a test: the consistency relation $\Ccal^2 = \Ccal^2(\alpha)$ of eq.~\eqref{eq:consistency_curve} is what makes the signature specific to stasis, and it requires the amplitude step at $\fend$ to be measurable independently. 

\begin{figure}[t]
  \centering
  \begin{subfigure}{0.49\linewidth}
    \centering
    \includegraphics[width=\linewidth]{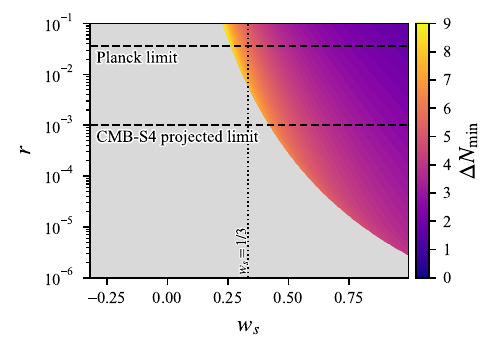}
    \caption{Band~I, $\fend = 10^{-4}\,\mathrm{Hz}$.}
    \label{fig:dnmin_f1}
  \end{subfigure}
  \hfill
  \begin{subfigure}{0.49\linewidth}
    \centering
    \includegraphics[width=\linewidth]{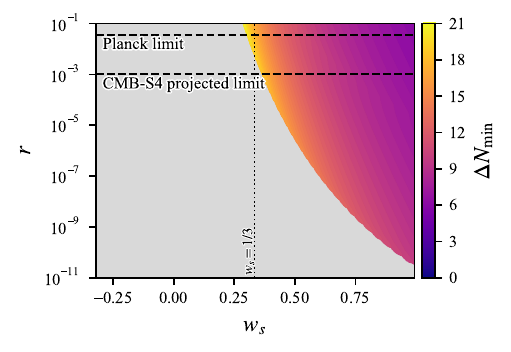}
    \caption{Band~II, $\fend = 10^{-9}\,\mathrm{Hz}$.}
    \label{fig:dnmin_f2}
  \end{subfigure}

  \vspace{0.5em}

  \begin{subfigure}{0.49\linewidth}
    \centering
    \includegraphics[width=\linewidth]{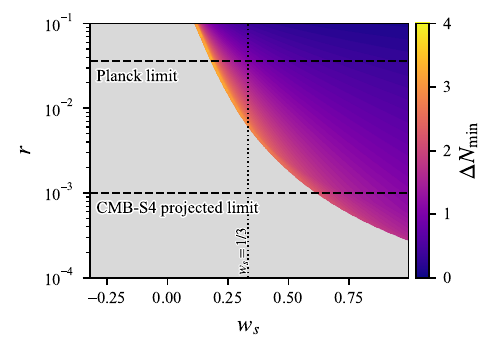}
    \caption{Band~III, $\fend = 10^{-2}\,\mathrm{Hz}$.}
    \label{fig:dnmin_f3}
  \end{subfigure}

  \caption{Shortest detectable stasis duration $\Delta N_\mathrm{min}$, in e-folds, as a function of the stasis equation of state $\ws$ and the tensor-to-scalar ratio $r$, for the three band anchors of Table~\ref{tab:bands}.  At each point the color gives eq.~\eqref{eq:dnmin} evaluated for the most sensitive instrument, using the PLS curves of~\protect\cite{Schmitz:2020rag} with $T_\mathrm{obs} = 4$\,yr and $\rho_\mathrm{thr} = 1$.  The detection criterion requires both the tilted stasis band and at least one adjacent plateau to cross that instrument's PLS curve, so that $\alpha$ and $\Ccal^2$ are separately measurable and the consistency relation of eq.~\eqref{eq:consistency_curve} can be tested.  Dashed lines: the Planck/BICEP-Keck upper limit $r < 0.036$~\protect\cite{BICEPKeck:2021gln} and the CMB-S4 projected limit $r < 10^{-3}$ at $95\%$~CL~\protect\cite{CMB-S4:2016ple}.  Grey: no stasis epoch within the plotted range of $\Delta N$ renders the feature detectable.  Note that $\fbeg$ is not fixed but follows from $(\ws, \Delta N)$, so the band labels denote the $\fend$ anchor only; the $\Delta N$ ranges, and hence the color scales, differ between panels.  The vertical dotted line marks $\ws = 1/3$.}
  \label{fig:dnmin}
\end{figure}

At fixed $\ws$ and $\fend$, $r_\mathrm{min}$ decreases monotonically with $\Delta N$ and then saturates, so it may be inverted to give the shortest stasis epoch that a given tensor amplitude renders detectable,

\begin{equation}
  \Delta N_\mathrm{min}(\ws, r) \;\equiv\; \min\bigl\{\Delta N \,:\, r_\mathrm{min}(\ws, \Delta N) \leq r \bigr\},
  \label{eq:dnmin}
\end{equation}

which is the quantity mapped in Fig.~\ref{fig:dnmin}. Each map fixes $\fend$ at one of the three band anchors of Table~\ref{tab:bands} and lets $\fbeg = \fend\exp[(1+3\ws)\Delta N/2]$ follow from $(\ws, \Delta N)$. We stress that the band labels are used here only as $\fend$ anchors, and that unlike the fixed-$(\fend,\fbeg)$ configurations of Table~\ref{tab:bands} the lever arm $\fbeg/\fend$ varies across each map, growing with both $\ws$ and $\Delta N$.  The $\ws$ axis spans $(-1/3, 1)$ continuously, so each map covers the full range of allowed equation of state values.  We take the PLS curves of~\cite{Schmitz:2020rag} scaled to $T_\mathrm{obs} = 4$\,yr and $\rho_\mathrm{thr} = 1$ throughout, and at each grid point evaluate eq.~\eqref{eq:dnmin} for whichever instrument performs best there.

At the Planck limit $r = 0.036$ the feature remains resolvable down to almost $\ws= 0.1$ in Band~III, and down to $\ws\approx 0.6$ at the CMB-S4 projected limit. The gray region marks parameter space in which the feature is undetectable for any $\Delta N$.

\begin{nota}
  Three caveats should be kept in mind for these plots. First, PLS curves are constructed for pure power-law backgrounds~\cite{Schmitz:2020rag}, so a broken power law that crosses one attains $\mathrm{SNR}\sim\rho_\mathrm{thr}$ only approximately; the maps are accurate at the level of the shape of the detectable region, not the precise placement of its boundary.  Second, resolving a deviation from a baseline formally calls for a model-comparison or Fisher analysis rather than an SNR threshold, and the criterion adopted here is the standard SNR-level proxy for it; a full Fisher treatment is deferred to future work (Sec.~\ref{sec:discussion}).  Third, the maps show the envelope over instruments and therefore identify where a detection is possible, but not by which instrument; the per-detector thresholds quoted in Secs.~\ref{sec:canonical} and ~\ref{sec:DS} and Appendices \ref{sec:MV} and \ref{sec:VR} are obtained from the same calculation applied to each PLS curve separately.
\end{nota}

\subsection{Microphysical Realizations of Stasis}

In this subsection we discuss the detectability of two specific realizations of stasis, canonical stasis in which the IGWB gets suppressed, and dynamical scalar stasis in which the IGWB gets enhanced. Discussion of the detectability of vacuum-energy/matter stasis and vacuum-energy/radiation stasis can be found in Appendices \ref{sec:MV} and \ref{sec:VR}.

\subsubsection{Canonical stasis \texorpdfstring{$(0\leq \ws < 1/3)$}{0<= ws < 1/3}}
\label{sec:canonical}

Canonical stasis~\cite{Dienes:2021woi} spans $\ws\in[0,1/3)$, giving
$\alpha\in(-2,0)$: the stasis band is suppressed relative to the RD baseline
(a notch), with suppression deepening as $\ws\to 0$ (matter domination).
The Bessel amplitude satisfies $\Ccal^2\in(9/16, 1)$, so the spectrum
steps down by at most $\sim 44\%$ at $\fend$.
The pre-stasis plateau is additionally suppressed by the accumulated tilt
factor $(\fbeg/\fend)^{\alpha}$, which can be orders of magnitude below the
RD baseline for long stasis durations.

Using power-law integrated sensitivity curves~\cite{Schmitz:2020rag} scaled
to $T_\mathrm{obs} = 4$\,yr and $\rho_\mathrm{thr} = 1$, BBO's minimum
sensitivity is $h^2\Omega_\mathrm{GW}^\mathrm{BBO}\approx 4\times 10^{-18}$,
just below the RD baseline of $6.8\times 10^{-18}$ at $r = 0.01$.  However,
the stasis feature is a \emph{suppression} of the IGWB, so detecting it
requires resolving the deviation below the baseline rather than the
baseline itself.  The suppression deepens as $\ws\to 0$, so the required tensor-to-scalar ratio grows monotonically with decreasing $\ws$. In Band III, the feature is detectable by BBO for $r\gtrsim 0.01$ at $\ws\approx 0.3$, rising to $r\gtrsim 0.036$ at $\ws\approx 0.2$. DECIGO achieves a minimum sensitivity of $\approx 1.2\times 10^{-17}$ (4\,yr), requiring $r \gtrsim 0.02$ for the flat baseline and somewhat
more for the suppressed stasis band.

For Band~I (LISA), neither the stasis band nor the RD baseline is reachable
at $r = 0.01$: LISA's minimum sensitivity $\sim 8 × 10^{-15}$ lies three orders
of magnitude above the IGWB for this value of $r$.  LISA becomes competitive
only for $r\gtrsim 10$, more than two orders of magnitude above current CMB bounds.

\begin{figure}[t]
  \centering
  \includegraphics[width=1\linewidth]{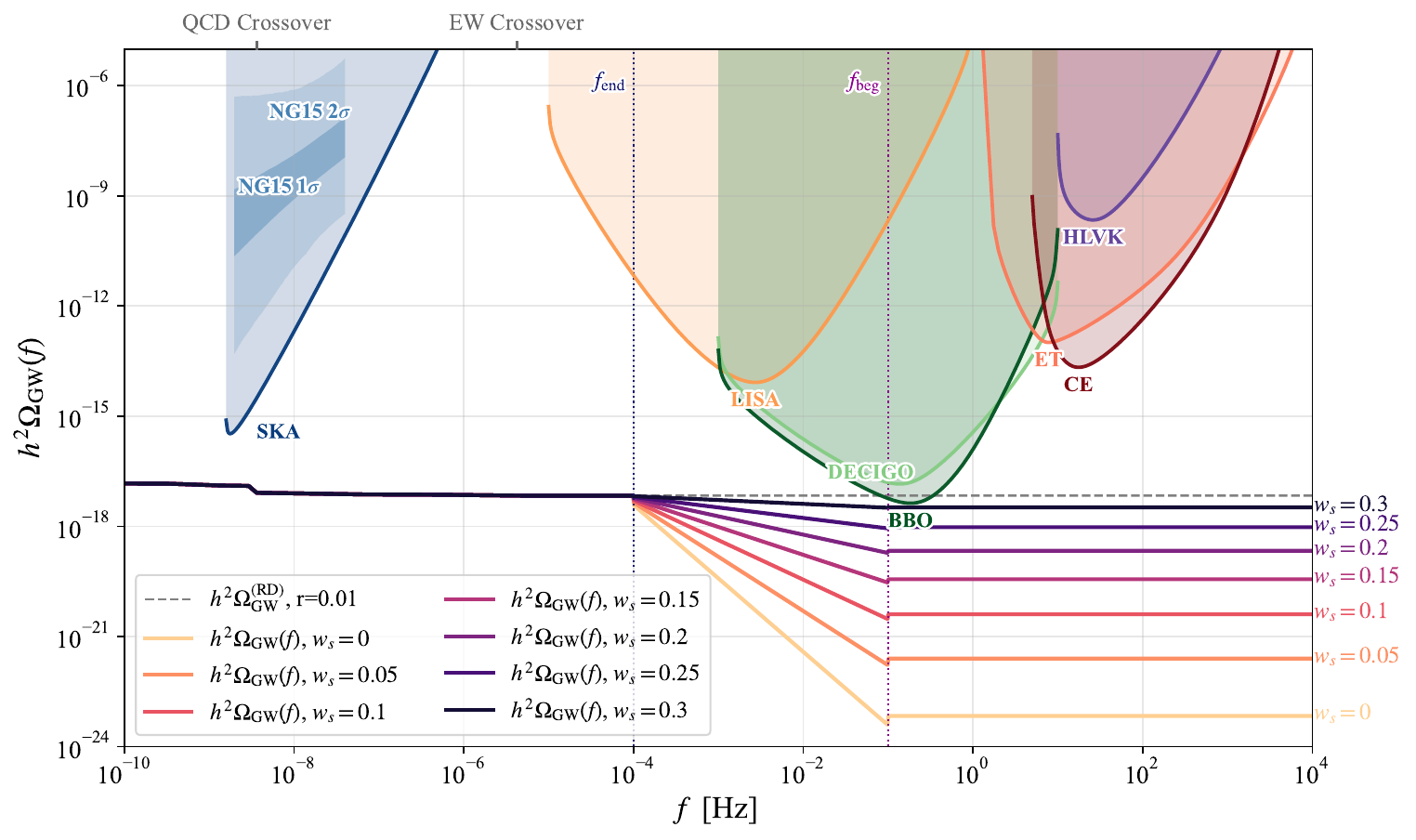}
  \caption{Stochastic GW background $h^2\Omega_{\rm GW}(f)$ for canonical
    stasis ($\ws\in[0,1/3]$) in Band~I
    ($\fend=10^{-4}\,\mathrm{Hz}$, $\fbeg=10^{-1}\,\mathrm{Hz}$;
    $T_\mathrm{end}\approx 3.8\,\mathrm{TeV}$).
    Grey dashed: RD baseline at $r=0.01$.
    Shaded: detector PLS curves~\protect\cite{Schmitz:2020rag} with $T_{obs}$ = 4 years and $\rho_{thr}$ = 1.
    Blue band: NANOGrav 15-year posterior~\protect\cite{NANOGrav:2023gor}.
    The feature falls in the LISA band; LISA's minimum sensitivity
    $\sim 8 × 10^-15$ requires $r\gtrsim 10^3$ for detection, four orders
    of magnitude above current Planck/BICEP/Keck bounds.}
  \label{fig:canon_f1}
\end{figure}

\begin{figure}[t]
  \centering
  \includegraphics[width=1\linewidth]{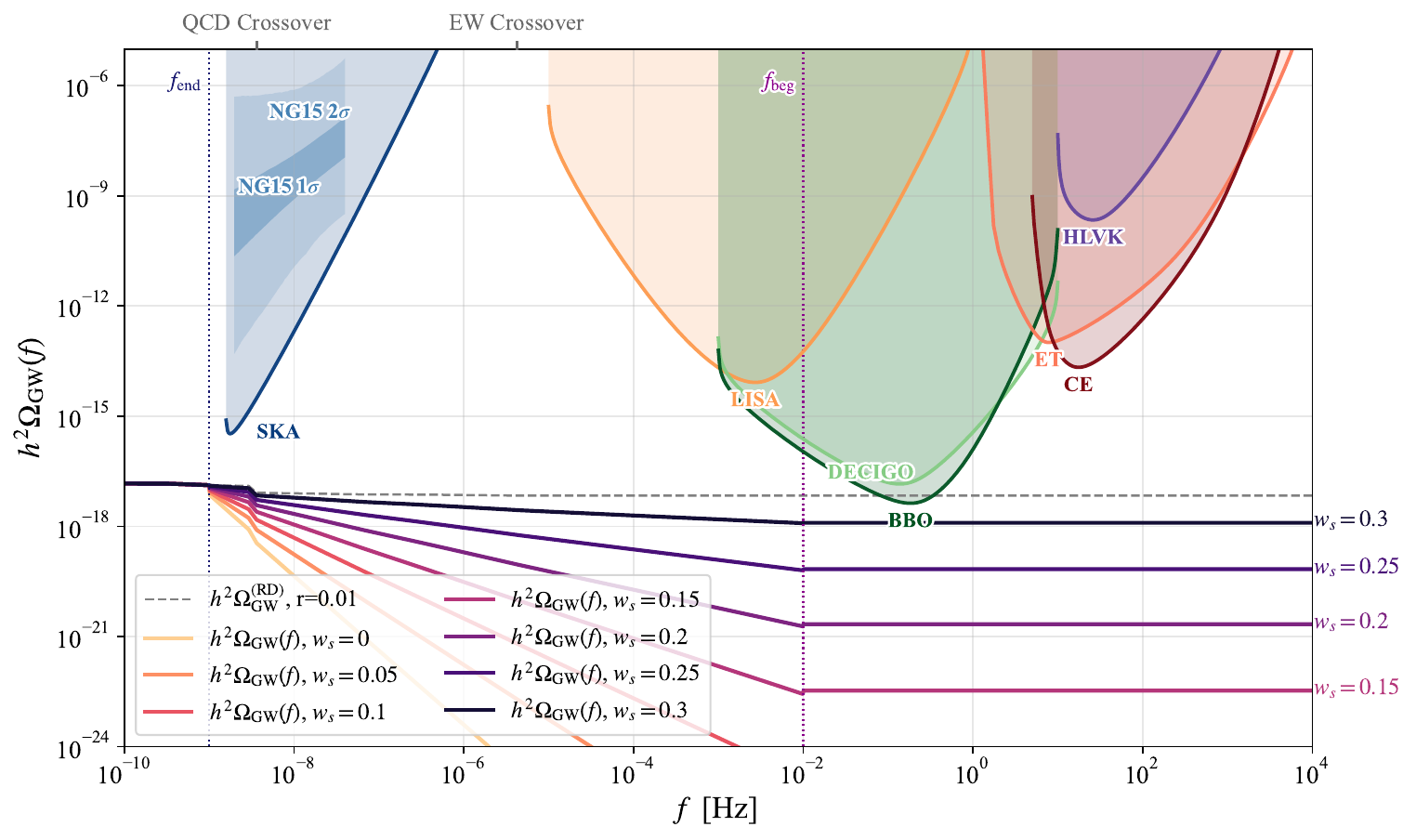}
  \caption{Same as Fig.~\ref{fig:canon_f1} for Band~II
    ($\fend=10^{-9}\,\mathrm{Hz}$, $\fbeg=10^{-2}\,\mathrm{Hz}$;
    $T_\mathrm{end}\approx 55\,\mathrm{MeV}$).
    The stasis band spans the NANOGrav-to-LISA range; the QCD and EW
    $g_*$ transitions at $3.6\times 10^{-9}$ and $2.6\times 10^{-6}\,\mathrm{Hz}$
    fall inside the stasis band (see Sec.~\ref{sec:gstar}).
    The end-of-stasis temperature is below the QCD crossover and
    near the BBN epoch, which imposes additional cosmological constraints
    on this scenario.}
  \label{fig:canon_f2}
\end{figure}

\begin{figure}[t]
  \centering
  \includegraphics[width=1\linewidth]{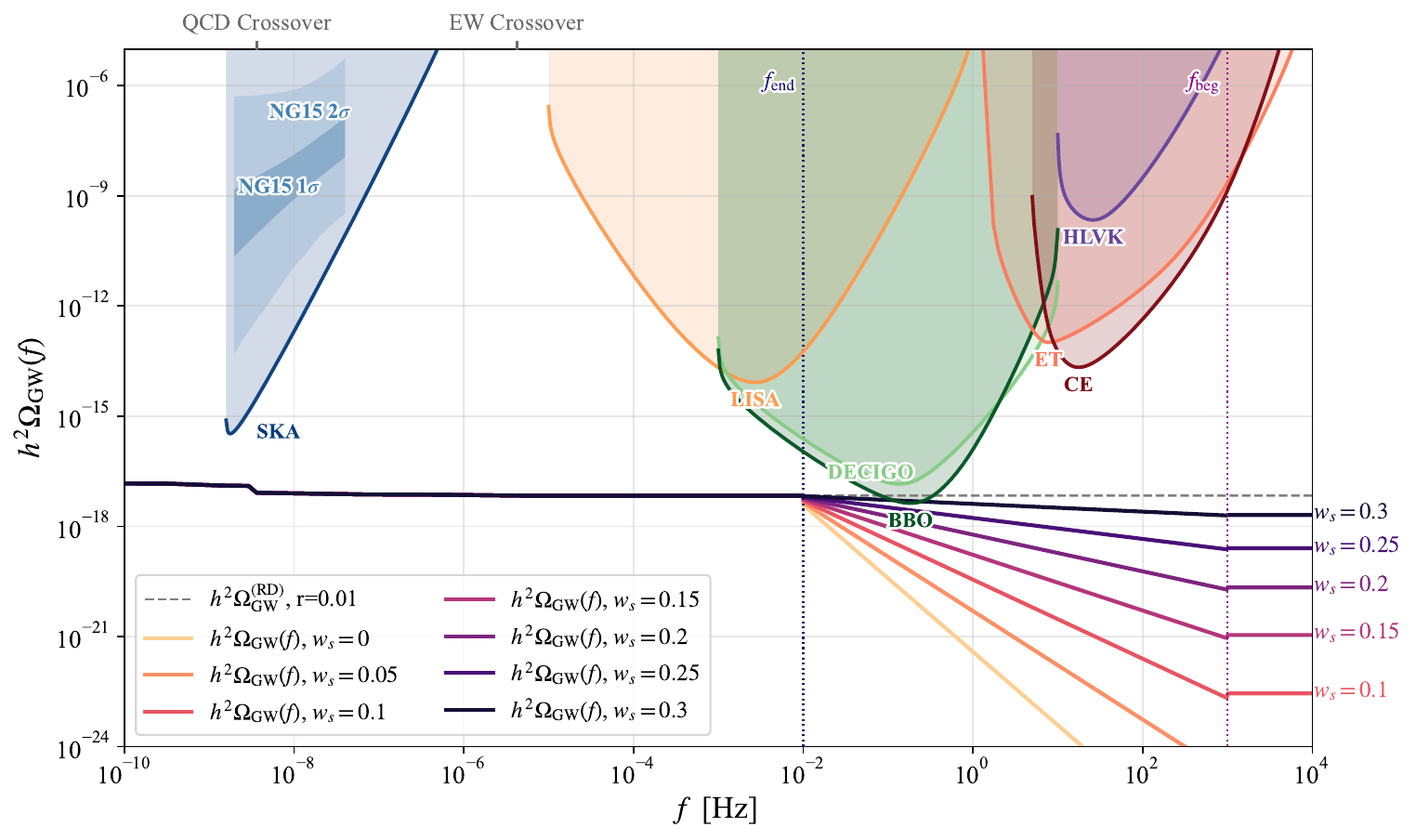}
  \caption{Same as Fig.~\ref{fig:canon_f1} for Band~III
    ($\fend=10^{-2}\,\mathrm{Hz}$, $\fbeg=10^{3}\,\mathrm{Hz}$;
    $T_\mathrm{end}\approx 380\,\mathrm{TeV}$).
    The stasis band falls squarely in the BBO/DECIGO-to-ET range.
    For $r\gtrsim\text{few}\times 0.01$ BBO
    can detect the stasis feature for large values of $\ws$ within the allowed range for canonical stasis.}
  \label{fig:canon_f3}
\end{figure}

\subsubsection{Dynamical scalar stasis \texorpdfstring{$(1/3 < \ws < 1)$}{1/3< ws < 1}}
\label{sec:DS}

\begin{figure}[t]
  \centering
  \includegraphics[width=1\linewidth]{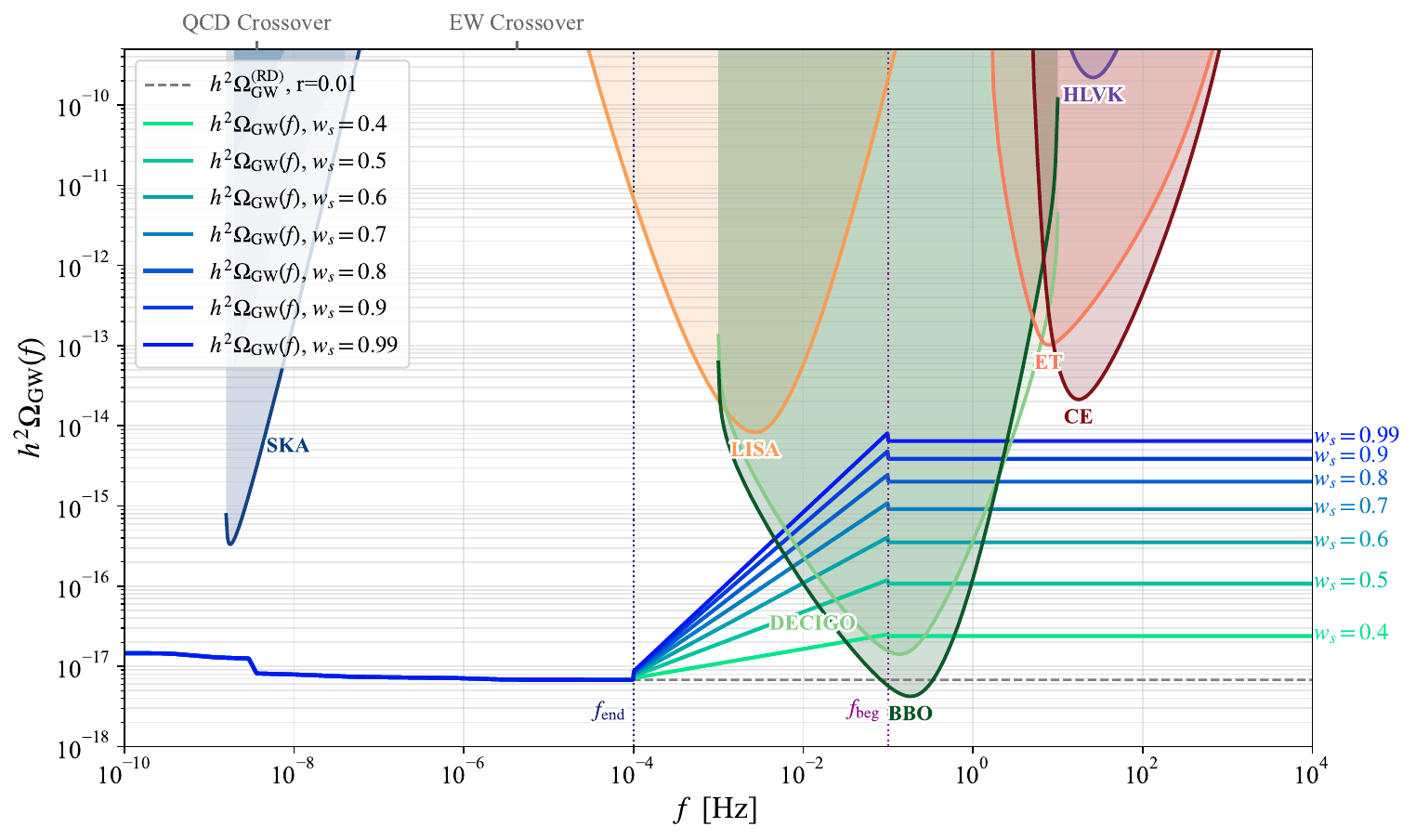}
  \caption{Stochastic GW background $h^2\Omega_{\rm GW}(f)$ for dynamical scalar
    stasis ($\ws\in[1/3,1]$) in Band~I
    ($\fend=10^{-4}\,\mathrm{Hz}$, $\fbeg=10^{-1}\,\mathrm{Hz}$;
    $T_\mathrm{end}\approx 3.8\,\mathrm{TeV}$).
    Grey dashed: RD baseline at $r=0.01$.
    Shaded: detector PLS curves~\protect\cite{Schmitz:2020rag} with $T_{obs}$ = 4 years and $\rho_{thr}$ = 1.
    Blue band: NANOGrav 15-year posterior~\protect\cite{NANOGrav:2023gor}.
    The feature falls in the LISA band; 
    $\sim 8 \times 10^{-15}$ requires $r\gtrsim 0.1$ for detection, above current Planck/BICEP/Keck bound of 0.036.}
  \label{fig:DS_f1}
\end{figure}

\begin{figure}[t]
  \centering
  \includegraphics[width=1\linewidth]{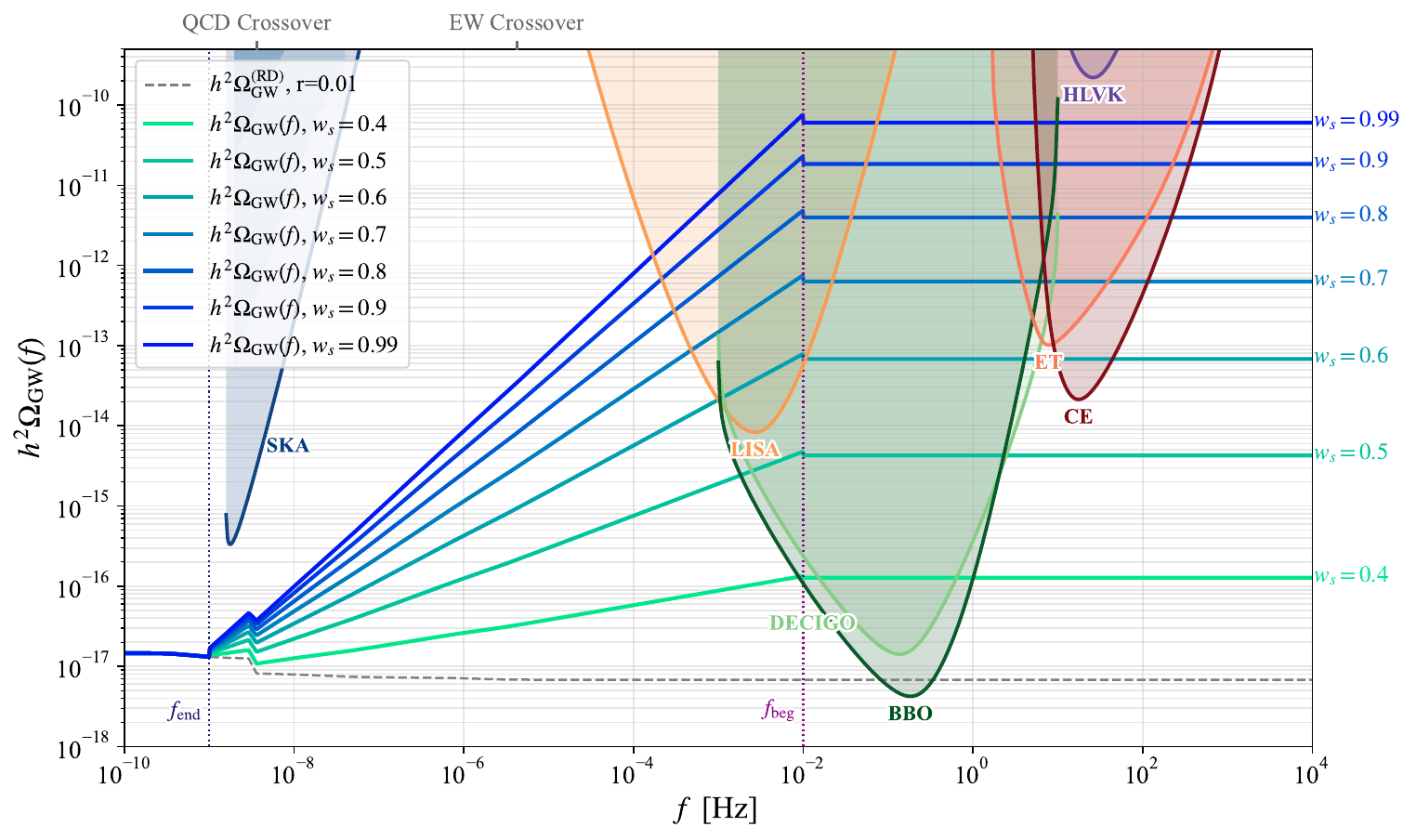}
  \caption{Same as Fig.~\ref{fig:DS_f1} for Band~II
    ($\fend=10^{-9}\,\mathrm{Hz}$, $\fbeg=10^{-2}\,\mathrm{Hz}$;
    $T_\mathrm{end}\approx 55\,\mathrm{MeV}$).
    The stasis band spans the NANOGrav-to-LISA range; the QCD and EW
    $g_*$ transitions at $3.6\times 10^{-9}$ and $2.6\times 10^{-6}\,\mathrm{Hz}$
    fall inside the stasis band (see Sec.~\ref{sec:gstar}).
    The end-of-stasis temperature is below the QCD crossover and
    near the BBN epoch, which imposes additional cosmological constraints
    on this scenario. BBO can detect the stasis feature across the entire dynamical scalar $\ws$ range at $r = 0.01$, with $r_\mathrm{min}$ falling below $10^{-7}$ for $\ws\gtrsim 0.8$ (see Sec.~\ref{sec:DS})}
  \label{fig:DS_f2}
\end{figure}

\begin{figure}[t]
  \centering
  \includegraphics[width=1\linewidth]{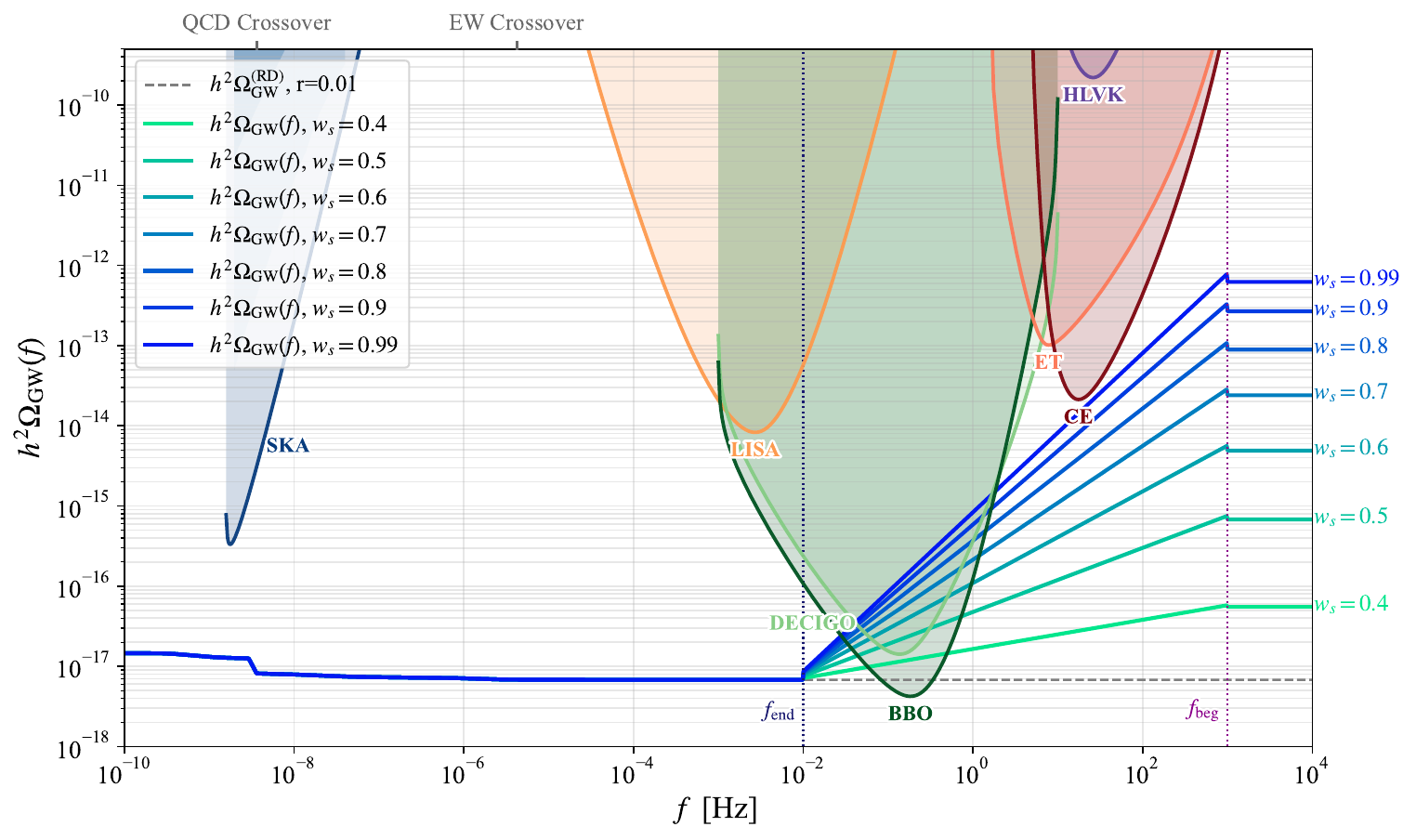}
  \caption{Same as Fig.~\ref{fig:DS_f1} for Band~III
    ($\fend=10^{-2}\,\mathrm{Hz}$, $\fbeg=10^{3}\,\mathrm{Hz}$;
    $T_\mathrm{end}\approx 380\,\mathrm{TeV}$).
    The stasis band falls squarely in the BBO/DECIGO-to-ET range.
    BBO can detect the stasis feature across the entire dynamical scalar $\ws$ range at $r = 0.01$.}
  \label{fig:DS_f3}
\end{figure}

\begin{figure}[t]
  \centering
  \includegraphics[width=1\linewidth]{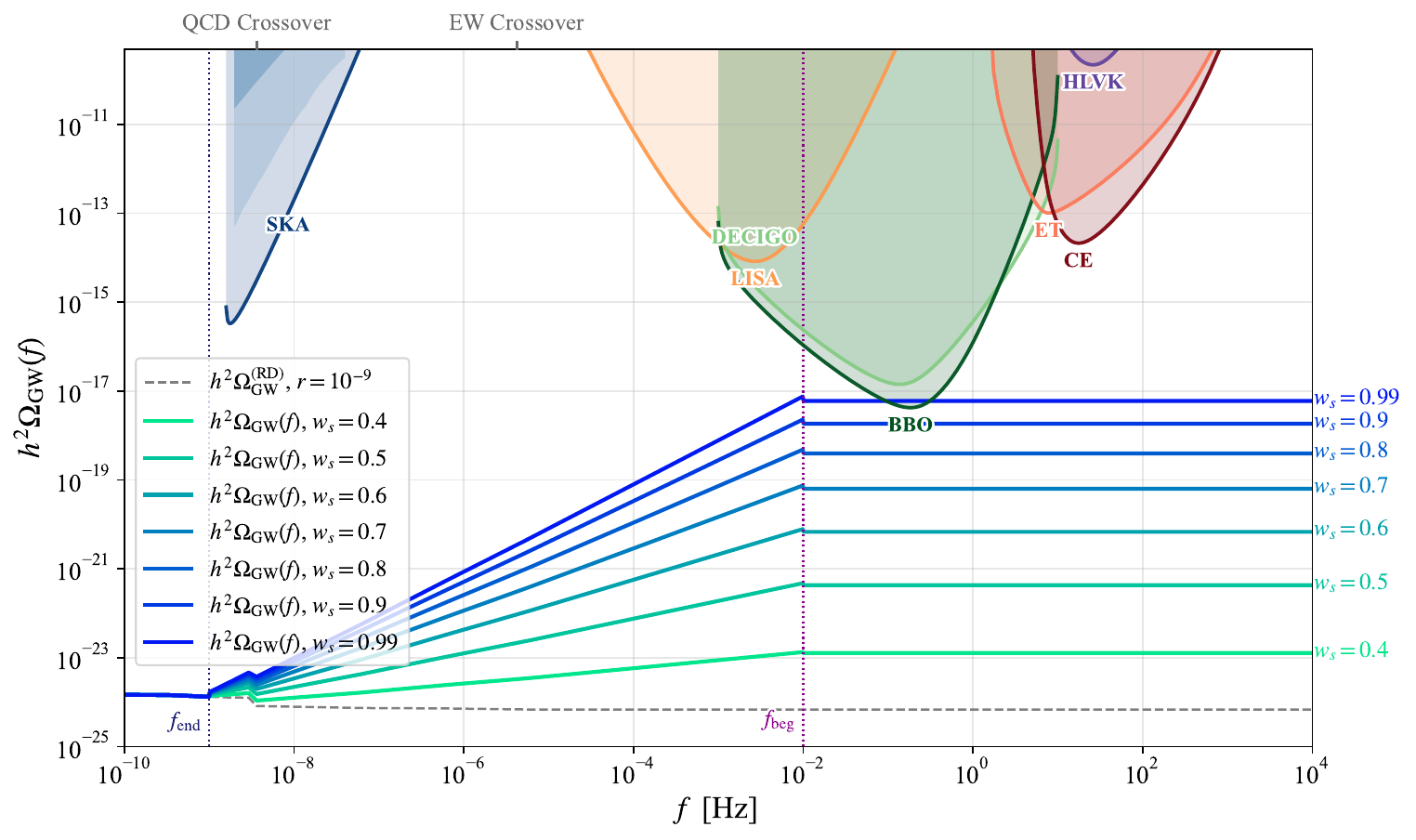}
  \caption{Stochastic GW background $h^2\Omega_{\rm GW}(f)$ for dynamical scalar
    stasis ($\ws\in[1/3,1]$) in Band~II with r set to the approximate minimum value for which the spectrum is still detectable by BBO for values of $w_s$ approaching the kination limit, $10^{-9}$;
    $T_\mathrm{end}\approx 55\,\mathrm{MeV}$.
    Grey dashed: RD baseline at $r=10^{-9}$.
    Shaded: detector PLS curves~\protect\cite{Schmitz:2020rag} with $T_{obs}$ = 4 years and $\rho_{thr}$ = 1.
    Blue band: NANOGrav 15-year posterior~\protect\cite{NANOGrav:2023gor}.}
  \label{fig:DS_f2_rsmall}
\end{figure}

Dynamical scalar stasis~\cite{Dienes:2024wnu} arises when a scalar field
sustains a fixed equation of state $\ws\in(1/3, 1)$ through a
potential-dominated attractor.  The key distinction from the canonical and
vacuum-energy scenarios is the sign of the spectral tilt: $\alpha(\ws) > 0$
for $\ws > 1/3$, so the stasis band is \emph{enhanced} relative to the
radiation-dominated baseline rather than suppressed.  The IGWB rises
through the stasis band from $\fend$ to $\fbeg$, reaching a pre-stasis
plateau elevated above the baseline by the accumulated-tilt factor
$(\fbeg/\fend)^\alpha$.  For Band~III with $\fbeg/\fend \sim 10^5$ and
$\ws \gtrsim 0.5$, this factor reaches $10^2$--$10^4$, and for $\ws$
approaching the kination limit ($\ws\to 1$, $\alpha\to 1$) it can exceed
$10^5$.  For Band~II, the plateau can be enhanced by as much as $\sim 10^7$. The crucial observational consequence is a decoupling from $r$:
even if $r$ is far below the projected sensitivity of any CMB experiment,
the stasis-enhanced plateau can still be visible to BBO and DECIGO. For Band~II, LISA and CE are competitive when r is of $O(0.01)$ for $w_s \gtrsim 0.55$ and ET is competitive when r is of $O(0.01)$ for $w_s \gtrsim 0.65$.
For all bands, BBO can detect the stasis feature across the entire
$\ws\in(1/3,1)$ parameter space at $r = 0.01$. As shown in Fig.\ref{fig:DS_f2_rsmall}, in Band~II $r_\mathrm{min}$
drops below $10^{-7}$ for $\ws \gtrsim 0.8$.  For $\ws$ close to the
kination limit, BBO can detect the feature for $r$ as small as
$\sim 10^{-9}$, well below the reach of CMB-S4 or LiteBIRD.
Figures~\ref{fig:DS_f1}--\ref{fig:DS_f3} show these spectra for the three
frequency bands.

In Bands I and II, both the spectral distortion, $\alpha$, and the amplitude step, $\Ccal^2$, are detectable for almost the full range of $w_s$ values. In Band III, however, only the enhancement is measurable, not the amplitude step. While in Bands I and II the consistency relation (Eq.~\ref{eq:consistency_curve}) can be used to verify the occurrence of a stasis epoch or one of its mimics, detection of the enhancement in Band III would at best be a hint of a stasis epoch.

\subsection{Consistency with the $\Delta N_\mathrm{eff}$ bound}
\label{sec:neff}

For the enhancement scenarios ($\ws > 1/3$), the pre-stasis plateau rises above the radiation-dominated baseline by the accumulated-tilt factor $(\fbeg/\fend)^\alpha$, which can reach $10^5$--$10^7$ for the longest-duration configurations. Because a gravitational-wave background contributes to the total radiation energy density, the integrated IGWB is bounded by the same big-bang-nucleosynthesis and CMB constraints on extra relativistic species that limit any dark-radiation component~\cite{Maggiore:1999vm, Caprini:2018mtu},
\begin{equation}
  \int h^2\Omega_\mathrm{GW}(f)\, d\ln f \;\lesssim\; 5.6\times 10^{-6}\,\Delta N_\mathrm{eff},
  \label{eq:neff_bound}
\end{equation}
where the integral runs over all modes that re-entered the horizon before the epoch at which the bound is applied. Taking the conservative post-Planck value $\Delta N_\mathrm{eff}\lesssim 0.2$ gives a ceiling of $\approx 1.1\times 10^{-6}$.

The integral is dominated by the high-frequency plateau, where $h^2\Omega_\mathrm{GW}$ is largest. For enhancement scenarios in which $\alpha > 0$ the plateau height is
\begin{equation}
  h^2\Omega_\mathrm{GW}^\mathrm{plateau} = \Ccal^2(\nu)\,\Bigl(\frac{\fbeg}{\fend}\Bigr)^{\!\alpha} h^2\Omega_\mathrm{GW}^{(\mathrm{RD})},
\end{equation}
and, since the integrand climbs monotonically through the stasis band and is flat above $\fbeg$, the integral is set by the plateau amplitude times its logarithmic extent, $\int h^2\Omega_\mathrm{GW}\,d\ln f \sim h^2\Omega_\mathrm{GW}^\mathrm{plateau}\,\Delta\ln f_\mathrm{plateau}$, with the stasis band itself contributing only an $\mathcal{O}(1/\alpha)$ correction.

The most stringent case is the one approaching kination ($\ws \rightarrow 1$, $\alpha \rightarrow 1$, $\Ccal^2 \rightarrow 4/\pi$) in Band~II, which combines the maximal tilt with the widest lever arm, $\fbeg/\fend = 10^7$. At the Planck upper limit $r = 0.036$ (so $h^2\Omega_\mathrm{GW}^{(\mathrm{RD})} = 2.4\times 10^{-17}$), the plateau reaches
\begin{equation}
  h^2\Omega_\mathrm{GW}^\mathrm{plateau} \approx \frac{4}{\pi}\times 10^{7}\times 2.4\times 10^{-17} \approx 3\times 10^{-10}.
\end{equation}
Even allowing the flat plateau to span a further $\sim 10$ decades in frequency above $\fbeg$, the integrated energy density is $\int h^2\Omega_\mathrm{GW}\,d\ln f \sim 10^{-9}$--$10^{-8}$, two to three orders of magnitude below the bound of Eq.~\eqref{eq:neff_bound}. All other enhancement configurations have either a smaller tilt, a shorter lever arm, or a lower $\fend$-to-$\fbeg$ ratio, and are correspondingly safer.

The enhancement scenarios therefore remain comfortably consistent with the $\Delta N_\mathrm{eff}$ bound across the entire $(\ws, \Delta N)$ parameter space considered here, even at the largest tensor-to-scalar ratios allowed by current data. Equivalently, Eq.~\eqref{eq:neff_bound} imposes an upper envelope on the combination $\Ccal^2(\fbeg/\fend)^\alpha\, r$; saturating it would require either a stasis duration well beyond $\Delta N \sim 40$ e-folds or a tensor amplitude far above the CMB limit, neither of which arises in the scenarios studied here.

\subsection{Which microphysical realizations land where?}
\label{sec:where}

The mapping from microphysics to detector band is set by the end-of-stasis
temperature $T_\mathrm{end}$ via eq.~\eqref{eq:fend}.  Different realizations
predict different $T_\mathrm{end}$ values and therefore probe different
frequency bands:
\begin{itemize}[leftmargin=1.5em]
  \item \textbf{Canonical tower / KK:} The decay rate of the last tower species
    is $\Gamma_\mathrm{last}\sim H(T_\mathrm{end})$, with $T_\mathrm{end}$
    set by the lightest species mass.  Towers with $m_\mathrm{min}\sim
    1\,\mathrm{TeV}$ give $T_\mathrm{end}\sim\mathrm{few}\,\mathrm{TeV}$
    and $\fend\sim 10^{-4}$--$10^{-3}\,\mathrm{Hz}$ (LISA band);
    lighter towers can push $T_\mathrm{end}$ down into the DECIGO/BBO band
    or below.
  \item \textbf{PBH-induced:} The end-of-stasis temperature is set by the
evaporation of the heaviest PBHs, with an evaporation temperature, $T_{\rm evap} \propto M_{\rm PBH}^{-3/2}$~\cite{Dienes:2022zgd}.  Light PBHs
($M \sim 10^4\,{\rm g}$) evaporate at $T \sim 10^4\,{\rm GeV}$, placing
$f_{\rm end}$ in the LISA band, while heavier PBHs move it down toward the PTA band.
  \item \textbf{Dynamical scalars:} The attractor exit temperature is set by
    the scalar potential; models explored in~\cite{Dienes:2024wnu} span a
    wide $T_\mathrm{end}$ range.
  \item \textbf{Thermal annihilation / field-dependent decay:} Both are
    sensitive to the specific coupling and field-dynamics parameters, making
    the frequency-band prediction realization-specific.
\end{itemize}

The detectability analyses of Sec.~\ref{sec:canonical} and Appendices \ref{sec:VR} and \ref{sec:MV}
show that BBO/DECIGO Band~III is the most promising configuration for suppression scenarios. This
places a rough requirement $T_\mathrm{end}\sim 10^5$--$4\times10^7\,\mathrm{GeV}$
on the preferred realization.  Canonical towers and some PBH scenarios satisfy this naturally; the field-dependent and gravitational-interaction realizations would need to be tuned to this range. For scenarios in which the spectrum is enhanced, the stasis feature may be detected in any of the three bands, but is most distinguishable from other models in Bands I and II as both $\alpha$ and $\Ccal^2$ can be detected.

\section{End-of-stasis transition: a phenomenological treatment}
\label{sec:transition}

The piecewise spectral template treats the transition out of stasis as
instantaneous: $\ws$ jumps discontinuously to $1/3$ at $a_\mathrm{end}$,
producing a sharp break in the spectrum at $\fend$, which is an approximation
and physically imprecise. The attractor that maintains constant $\Omr$ requires
a continuous source of injected energy, and that source must shut off over some
finite window: the longest-lived species in a decaying tower has a finite decay
time, the lightest PBHs in a mass spectrum evaporate over a tail, the dynamical
scalar transitions out of the slow-roll-like regime over a finite range of
$\phi$, and so on. The result is a smooth transition in $w(t)$ between $\ws$
and $1/3$ over some number of e-folds, and a corresponding smoothing of the
spectral break.

The same picture applies to the vacuum-energy/matter and vacuum-energy/radiation
stases of~\cite{Dienes:2023ziv}: the attractor is sustained by a tower of
vacuum-like species with equation of state $w$, and the epoch ends when the
longest-lived member of that tower depletes. The transition width is again set
by the lifetime of this last component relative to $H^{-1}$ at $a_\mathrm{end}$,
giving $\Delta N_\mathrm{trans} = \mathcal{O}(1)$ e-folds. Throughout this section we assume the epoch in question lies
inside the validity range $\ws > -1/3$; for accelerating epochs, the
inflation-like treatment replaces the smoothed-break picture entirely, since
there is no in-epoch horizon-crossing event to smooth.

Computing $w(t)$ exactly through the transition is mechanism-specific and is
outside of the scope of this paper. The strategy here is instead
phenomenological: characterize the transition by a single width parameter and
argue analytically how that width smooths the spectrum.

\subsection{The transition width $\Delta N_\mathrm{trans}$}
\label{sec:DeltaNtrans}

We define $\Delta N_\mathrm{trans}$ as the number of e-folds over which the
total equation of state interpolates between $\ws$ (the attractor value) and
$1/3$ (post-stasis RD) at the end of stasis:
\begin{equation}
  \Delta N_\mathrm{trans}
  \;\equiv\; \ln\!\left(\frac{a_\mathrm{trans, end}}{a_\mathrm{trans, beg}}\right),
  \label{eq:DeltaNtrans}
\end{equation}
where $a_\mathrm{trans, beg}$ and $a_\mathrm{trans, end}$ bracket the window
over which $w$ moves substantially from $\ws$ to its RD value, $1/3$.

The onset of stasis at $\fbeg$ is characterised by an independent width
$\Delta N_\mathrm{trans}^\mathrm{(onset)}$, which is generically different from
$\Delta N_\mathrm{trans}$ because attractor establishment and dissolution are
governed by different physics.

\paragraph{Mechanism-dependent expectations.}
\begin{itemize}[leftmargin=1.5em]
  \item \textbf{Decaying tower (canonical).} The attractor dissolves over a
    few decay times of the longest-lived species, giving $\Delta
    N_\mathrm{trans} = \mathcal{O}(1)$ e-folds.

  \item \textbf{PBH-induced.} The shape of the lower edge of the PBH mass
    spectrum determines the transition width: narrow spectra give sharper
    transitions, broad spectra wider ones.

  \item \textbf{Vacuum-energy stases.} The same
    $\mathcal{O}(1)$ e-fold expectation as the canonical case, with the
    precise value set by the vacuum-energy tower structure.

  \item \textbf{Dynamical scalar.} The transition width is controlled by the
    shape of $V(\phi)$ near the exit: steep potentials give sharp transitions,
    flat ones gradual ones. No general estimate is possible.
\end{itemize}

The takeaway is that $\Delta N_\mathrm{trans} = \mathcal{O}(1)$ e-folds is
the natural expectation for all realizations with a clear depletion timescale,
while mechanism-specific cases require dedicated analysis.

\subsection{How the transition smooths the spectrum}
\label{sec:smoothing}

Modes deep inside the horizon during the transition redshift as $h_k \propto
1/a$ regardless of the background equation of state. Modes whose horizon
crossing falls within the transition window are affected by the changing
equation of state, and their amplitude is interpolated between the stasis-side
value $\Ccal^2(\nu_s)$ and the RD-side value $\Ccal^2(1/2) = 1$.

The relation between e-folds of expansion and log-frequency of
horizon-crossing modes is set by $aH \propto a^{-(1+3w)/2}$, so a mode
crossing $\Delta N$ e-folds later sits at log-frequency
\begin{equation}
  \Delta \ln f = \frac{1+3w}{2}\,\Delta N
  \label{eq:dlnf_dN}
\end{equation}
relative to the earlier-crossing mode. The total log-frequency range of modes
crossing during the transition is therefore
\begin{equation}
  \Delta \ln f_\mathrm{smooth}
  = \int_{N_\mathrm{trans,beg}}^{N_\mathrm{trans,end}}
    \frac{1+3w(N)}{2}\,dN,
  \label{eq:smoothing_width}
\end{equation}
which is bounded between
\begin{equation}
  \frac{1+3\ws}{2}\,\Delta N_\mathrm{trans}
  \;\leq\; \Delta \ln f_\mathrm{smooth}
  \;\leq\; \Delta N_\mathrm{trans}.
  \label{eq:smoothing_bounds}
\end{equation}
For any monotonic profile $w(N)$ that is symmetric about its midpoint (linear,
tanh, cubic smoothstep, and most other natural choices), the integral has the closed
form
\begin{equation}
  \Delta \ln f_\mathrm{smooth}^{(\mathrm{sym})}
  = \frac{3(1+\ws)}{4}\,\Delta N_\mathrm{trans}
  = \frac{1+3\bar{w}}{2}\,\Delta N_\mathrm{trans},
  \qquad
  \bar{w} \equiv \frac{\ws + 1/3}{2},
  \label{eq:smoothing_linear}
\end{equation}
as the integral depends only on the average of $w$ over the
transition, equal to $\bar w$ for any symmetric profile. The result sits at
the \emph{arithmetic} midpoint of the bounds in eq.~\eqref{eq:smoothing_bounds}
and serves as a representative value for the $\mathcal{O}(1)$ prefactor. Note that eq.~\eqref{eq:smoothing_bounds} is written for the suppression
regime $\ws < 1/3$, in which $(1+3\ws)/2 < 1$; for enhancement scenarios
with $\ws > 1/3$ the prefactor exceeds unity and the two bounds exchange
roles, so that $\Delta N_\mathrm{trans}$ becomes the lower bound and
$(1+3\ws)\Delta N_\mathrm{trans}/2$ the upper.

\begin{nota}
  Eq.~\eqref{eq:smoothing_width} is the transition analogue of the
  stasis-band width equation.  The latter is a special case in which
  $w(N) = \ws$ is constant, giving $\Delta\ln f = (1+3\ws)/2\cdot\Delta N$.
  The transition integrates the same kinematic relation over a window in
  which $w$ is varying. No new physics is needed.
\end{nota}

\subsubsection*{Deep sub-horizon modes are unaffected}

The tensor mode equation $\mu_k'' + (k^2 - a''/a)\mu_k = 0$ in the WKB
limit $k^2\gg a''/a$ gives $\mu_k\approx C\cos(k\tau+\phi)/\sqrt{k}$, a
pure oscillation whose amplitude $C$ is frozen at horizon crossing and
conserved thereafter, independent of the subsequent background dynamics. Any
mode that crossed the horizon before the transition has its amplitude locked in
by the stasis-era Bessel matching; the transition merely rescales it by $1/a$.
These modes contribute to the pre-stasis plateau with the standard accumulated
tilt $(\fbeg/\fend)^\alpha$, independent of $\Delta N_\mathrm{trans}$.
Modes that have not yet crossed by $a_\mathrm{trans,end}$ will cross in the
post-stasis RD era and pick up $\Ccal^2(1/2) = 1$, with no memory of the
transition shape.

\subsubsection*{Sharp transitions and Bogoliubov oscillations}

The smooth interpolation above is valid when the adiabaticity condition
$\Delta N_\mathrm{trans}\gg 1$ holds.  When $\Delta N_\mathrm{trans}\lesssim 1$,
the transition is sudden relative to the mode's oscillation period at horizon
crossing, and the Bessel matching can produce non-trivial Bogoliubov mixing:
small oscillations imprinted on the spectrum near $\fend$. Computing their
amplitude and frequency requires solving the mode equation through a specific
$w(N)$ profile and is left for future work.

\subsection{Consequences for future observation}
\label{sec:smoothing_observables}

The smoothing does not break the consistency framework but changes how the
observables are extracted. The slope $\alpha$ is robust to the smoothing
provided that $\Delta N_\mathrm{stasis}\gg \Delta N_\mathrm{trans} + \Delta
N_\mathrm{trans}^\mathrm{(onset)}$: in this regime the shoulders at $\fend$
and $\fbeg$ affect only a small fraction of the stasis band, leaving a clean
log-linear interior for the slope fit. The amplitude step $\Ccal^2$ becomes a
plateau-ratio fit rather than a measured jump: it is extracted from the
asymptotic ratio between the post-stasis plateau (extrapolated through the
shoulder) and the stasis-band power law.  The consistency relation $\Ccal^2 =
\Ccal^2(\alpha)$ from~\paperI~continues to hold in the smoothed template.

\begin{nota}
  The bandwidth condition $\Delta N_\mathrm{stasis}\gg \Delta N_\mathrm{trans}
  + \Delta N_\mathrm{trans}^\mathrm{(onset)}$ is the practical criterion for
  whether the stasis feature is observationally well-resolved.  For canonical
  $\Delta N_\mathrm{trans} = \mathcal{O}(1)$, this requires $\Delta
  N_\mathrm{stasis}\gtrsim$ a few, satisfied by all scenarios of
  phenomenological interest.
\end{nota}

\begin{figure}[t]
  \centering
  \includegraphics[width=\textwidth]{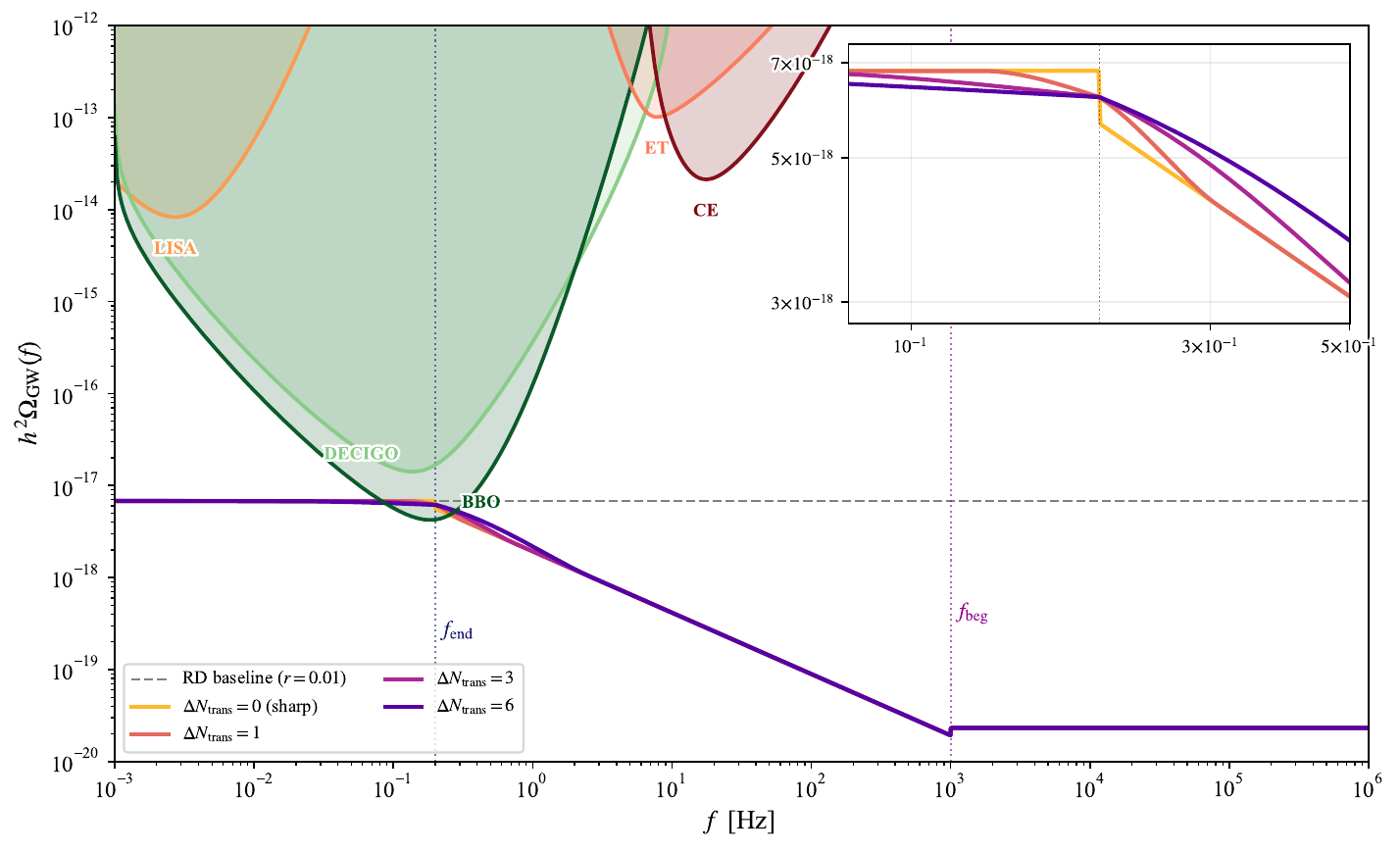}
  \caption{Spectral template at canonical stasis ($\ws=1/6$, $\alpha=-2/3$)
    with the lower break at $\fend=2\times10^{-1}\,\mathrm{Hz}$ smoothed over the
    predicted width of eq.~\eqref{eq:smoothing_linear}, for $\Delta
    N_\mathrm{trans}\in\{0,1,3,6\}$. The upper break at $\fbeg$ is kept
    sharp. The inset zooms in on the shoulder at $\fend$: $\Delta
    N_\mathrm{trans}=1$ produces a small rounding, while $\Delta
    N_\mathrm{trans}=6$ broadens the shoulder over more than a decade. The
    post-stasis plateau and the stasis-band power law are unaffected away
    from $\fend$. Shaded regions: PLS curves, 4 year observation time~\cite{Schmitz:2020rag}; gray
    dashed: RD baseline at $r=0.01$.}
  \label{fig:smoothing}
\end{figure}

Figure~\ref{fig:smoothing} shows the spectral template at canonical stasis
($\ws = 1/6$, $\alpha = -2/3$) with the lower break smoothed over the
predicted width of eq.~\eqref{eq:smoothing_linear}, for $\Delta N_\mathrm{trans}
\in \{0, 1, 3, 6\}$, with $\fend = 10^{-1}\,\mathrm{Hz}$ placing the
smoothed shoulder in the BBO/DECIGO band.

\section{Discussion and outlook}
\label{sec:discussion}

We have presented the full observational prediction for the stasis imprint on
the inflationary gravitational wave background, mapped onto the sensitivity
landscape of current and future detectors.  The main conclusions are:

\paragraph{BBO and DECIGO are the primary instruments.}
Among all planned detectors, BBO and DECIGO offer the most favorable
combination of sensitivity and frequency coverage for detecting the stasis
feature in the IGWB, with the relative prospects depending strongly on
whether the stasis band produces a suppression ($\ws < 1/3$) or an
enhancement ($\ws > 1/3$).

For \emph{suppression scenarios} ($\ws < 1/3$), the stasis feature is
a notch below the RD baseline.  Using the Schmitz~\cite{Schmitz:2020rag}
power-law integrated sensitivity curves scaled to $T_\mathrm{obs} = 4$\,yr
and $\rho_\mathrm{thr} = 1$, BBO can detect the stasis feature in Band~III
for $r \gtrsim 0.01$--$0.05$ depending on $\ws$, with deeper notches
(smaller $\ws$) requiring larger $r$.  DECIGO requires $r \gtrsim 0.02$
for a flat spectrum and somewhat more for the suppressed stasis band.
LISA, ET, and CE are competitive only for $r\gg 1$, far above current
CMB bounds.

For \emph{enhancement scenarios} ($\ws > 1/3$), the pre-stasis plateau
rises above the baseline by factors of $10^2$--$10^5$ depending on $\ws$
and the stasis duration.  Both BBO and DECIGO can detect the enhancement
feature across the entire $(\ws, \Delta N)$ parameter space at $r = 0.01$
for Band~III. For $\ws$ approaching the kination limit, BBO can detect
the feature at $r$ as small as $10^{-9}$ for Band~II.  This decouples the
stasis GW signature from the tensor amplitude entirely: a universe with
unmeasurably small $r$ can still leave a detectable stasis imprint in the
BBO band if the dynamical scalar equation of state is sufficiently large.
The combined reach of BBO and DECIGO established in~\paperI~further
strengthens this conclusion: jointly, the two instruments can resolve the
consistency relation $\Ccal^2 = \Ccal^2(\alpha)$ to $\sigma_\perp \simeq
1.5\times10^{-5}$ at $r = 0.01$, providing a sharp test of the stasis
hypothesis independent of the underlying microphysical realization.

\paragraph{Three independent observables.}
The stasis GW signature provides three physically distinct observables:
(i) the spectral tilt $\alpha$ in the stasis band, which directly encodes $\ws$;
(ii) the amplitude step $\Ccal^2$ at $\fend$, which provides the
  internal consistency check $\Ccal^2 = \Ccal^2(\alpha)$ established in~\paperI;
(iii) the break-shoulder width $\Delta\ln f_\mathrm{smooth}$, which encodes
  the transition microphysics through $\Delta N_\mathrm{trans}$.
A detection that simultaneously measures all three quantities would give
$\ws$, $T_\mathrm{end}$, $\Delta N_\mathrm{stasis}$, and $\Delta N_\mathrm{trans}$:
a remarkably complete characterization of the stasis epoch.

\paragraph{The $g_*$ fine structure as a cross-check.}
For Band~II scenarios ($T_\mathrm{end}\lesssim T_\mathrm{QCD}$), the Standard
Model $g_*$ transitions at $f_\mathrm{QCD}$ and $f_\mathrm{EW}$ produce
spectral steps at known, fixed frequencies.  These steps must be correctly
modeled to extract $\alpha$ and $\Ccal^2$ without bias, but simultaneously
provide a non-trivial cross-check: a detection of stasis in Band~II should
reproduce both the stasis-induced breaks at $\fend, \fbeg$ \emph{and} the
SM-predicted steps at $f_\mathrm{QCD}, f_\mathrm{EW}$, with the latter at
the correct amplitudes. Agreement would constitute strong evidence that the
observed spectral feature is genuinely cosmological.

\paragraph{Future directions.}
Several items are deferred to future work.  A full Fisher-matrix analysis
(parameter estimation precision for $\ws$, $\Delta N$, $T_\mathrm{end}$, $r$)
using realistic BBO/DECIGO noise models and including the $g_*$ corrections
as fixed nuisance parameters would quantify the measurement reach more
precisely than the SNR-based analysis of~\paperI.  The $g_*$ corrections
for Band~II also motivate a dedicated study of stasis scenarios with
$T_\mathrm{end}$ near the QCD scale, where the interplay between the stasis
spectrum and the hadronic transition could produce richer spectral features.
Finally, computing $\Delta N_\mathrm{trans}$ from first principles for specific
realizations, particularly the PBH case~\cite{Dienes:2022zgd} and the
canonical tower~\cite{Dienes:2021woi}, would turn the phenomenological
treatment of Sec.~\ref{sec:transition} into quantitative predictions.

\section*{Acknowledgments}

We would like to thank Tim Tait for the support and valuable feedback on our work. GB is supported by the Spanish grant  PID2023-151418NB-I00 funded by MCIU/AEI/10.13039/501100011033. AKB acknowledges support from the ``Unit of Excellence Maria de Maeztu 2020-2023'' award to the ICC-UB CEX2019-000918-M and grant PID2022-136224NB-C21 funded by MCIN / MINECO / MCOC. AKB and GB are supported by the European Union’s Horizon 2020 research and innovation program under the Marie Skłodowska-Curie grant agreement No 860881-HIDDeN, and Horizon Europe research and innovation program under the Marie Skłodowska-Curie Staff Exchange grant agreement No 101086085 – ASYMMETRY. 

\bibliographystyle{apsrev4-2}
\bibliography{biblio}

\appendix

\section{Vacuum-energy/matter stasis (\texorpdfstring{$-1/3 < \ws < 0$}{-1/3 < ws < 0})} 
\label{sec:MV}

In the vacuum-energy/matter scenario of~\cite{Dienes:2023ziv} (Sec.~III therein),
the stasis attractor sustains a non-zero vacuum-energy fraction alongside matter,
giving an effective equation of state $\ws\in(-1/3, 0)$.  The spectral tilt
$\alpha < -2$ is more negative than in the canonical case, meaning the stasis
band is \emph{more} strongly suppressed than under matter domination.
The Bessel amplitude $\Ccal^2 < 9/16$ decreases further below unity as $\ws\to -1/3$, the validity boundary below which the expansion is accelerating and no
in-stasis horizon crossing occurs.

This additional suppression has two observational consequences.
First, for fixed $r$ the signal in the stasis band is weaker than in the
canonical case, making detection harder.  Second, the pre-stasis plateau
is suppressed by $(\fbeg/\fend)^{\alpha}$ with $\alpha < -2$, which for
moderate $\Delta N$ already brings the plateau to effectively zero, making
the multi-epoch comb structure discussed in~\paperI~correspondingly harder to resolve.

\begin{figure}[t]
  \centering
  \includegraphics[width=1\linewidth]{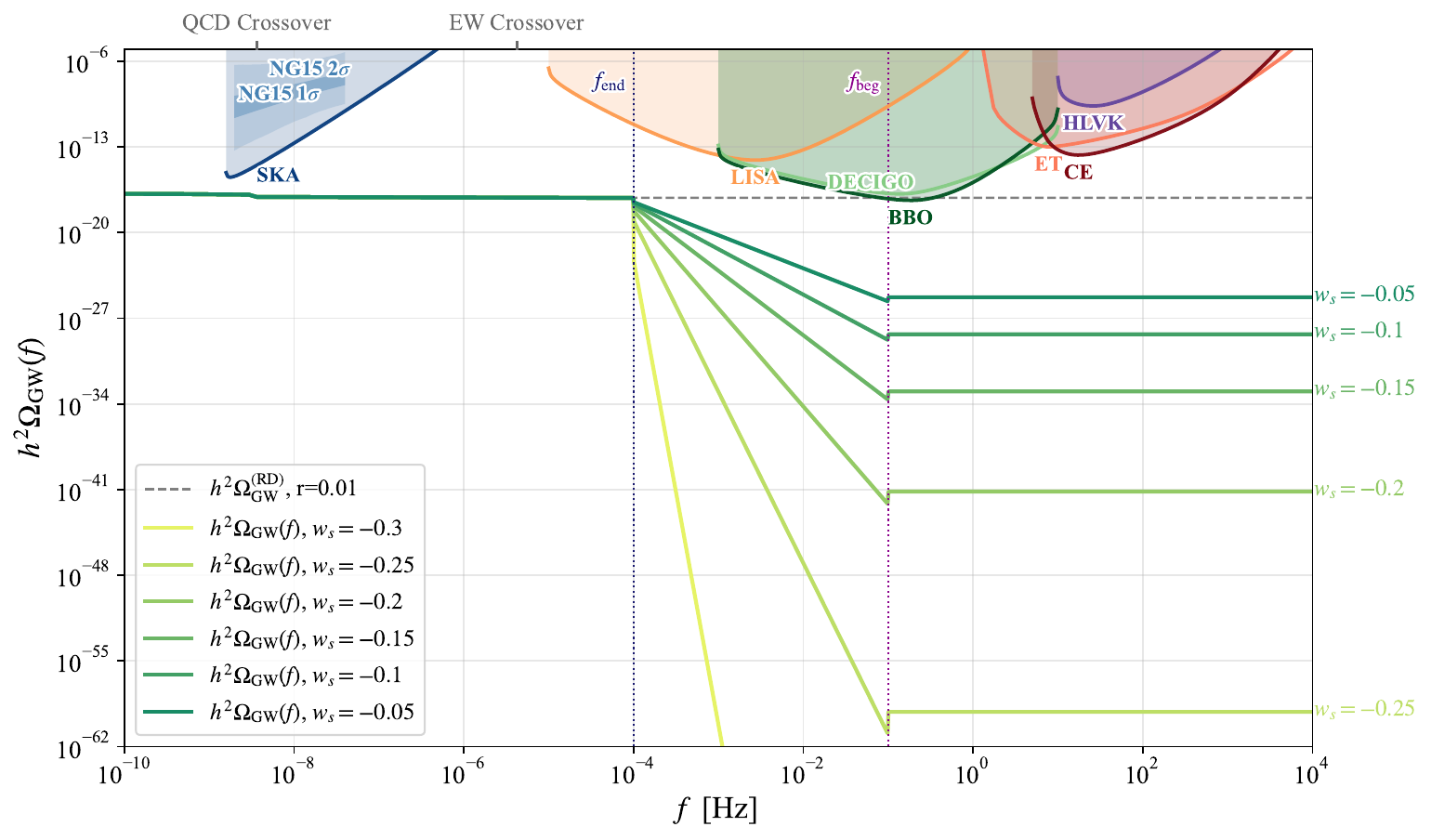}
  \caption{Stochastic GW background $h^2\Omega_{\rm GW}(f)$ for
    vacuum-energy/matter stasis ($\ws\in(-1/3, 0)$) in Band~I.
    Grey dashed: RD baseline at $r=0.01$.
    Shaded: PLS curves~\protect\cite{Schmitz:2020rag}.
    Blue band: NANOGrav 15-year posterior~\protect\cite{NANOGrav:2023gor}.
    The deeper suppression relative to canonical stasis is visible as a
    more negative slope $\alpha$ in the stasis band.}
  \label{fig:MV_f1}
\end{figure}

\begin{figure}[t]
  \centering
  \includegraphics[width=1\linewidth]{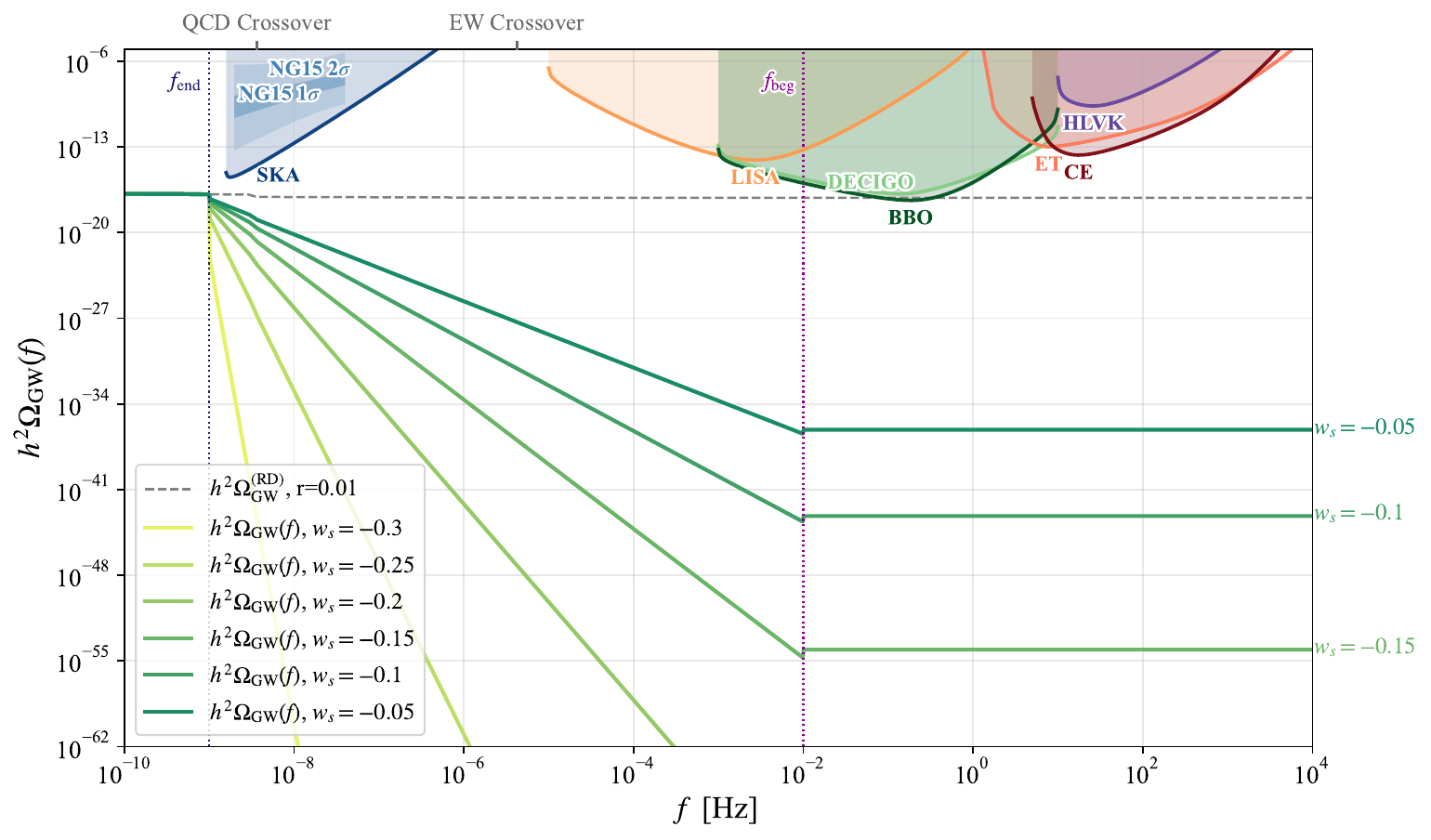}
  \caption{Same as Fig.~\ref{fig:MV_f1} for Band~II
    ($T_\mathrm{end}\approx 55\,\mathrm{MeV}$).  The $g_*$ steps at
    $f_\mathrm{QCD}$ and $f_\mathrm{EW}$ fall within the stasis band
    for this configuration.}
  \label{fig:MV_f2}
\end{figure}

\begin{figure}[t]
  \centering
  \includegraphics[width=1\linewidth]{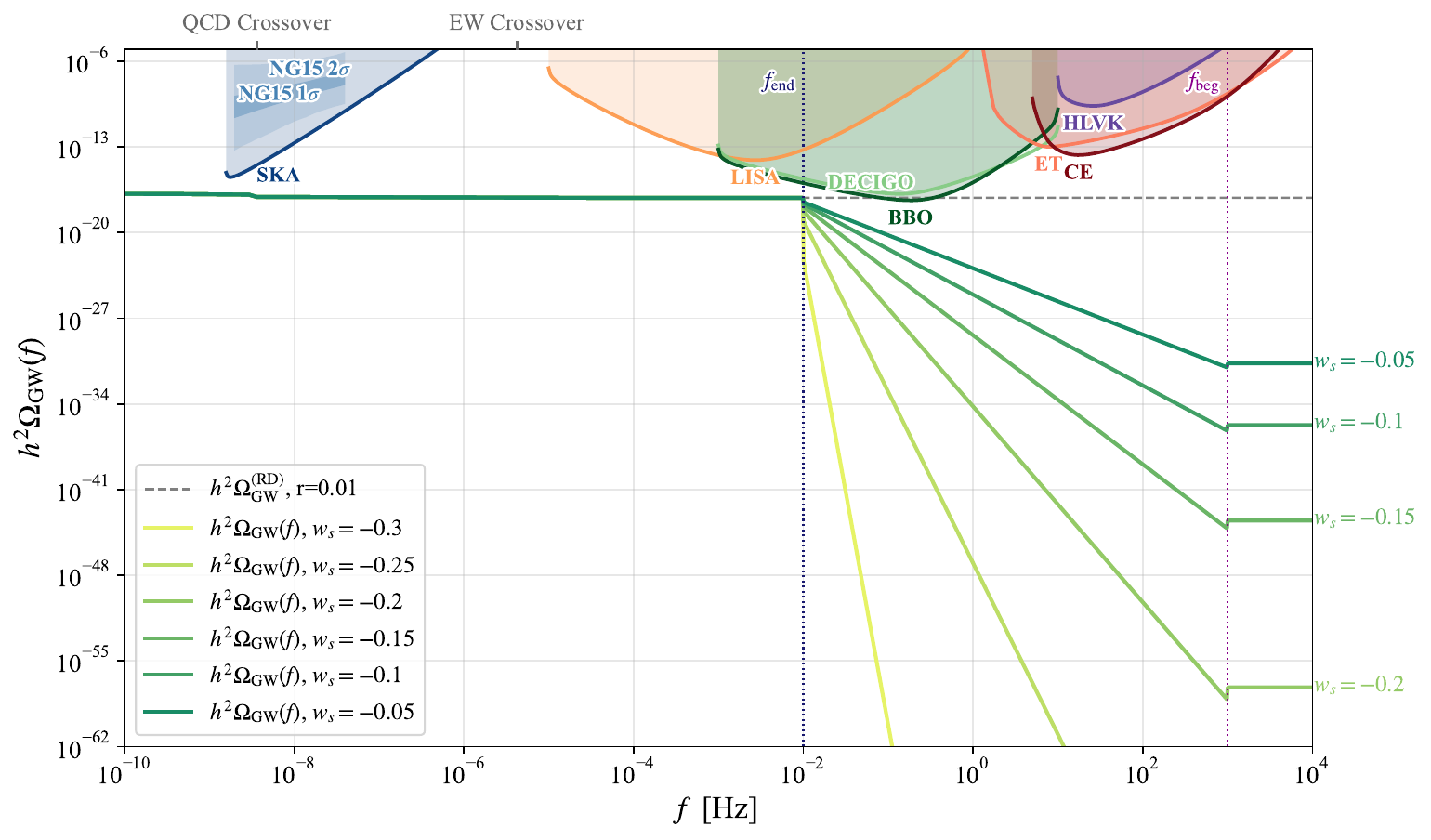}
  \caption{Same as Fig.~\ref{fig:MV_f1} for Band~III.
    The steep suppression at high frequencies makes detection
    progressively harder for $\ws\ll 0$.}
  \label{fig:MV_f3}
\end{figure}

Figures~\ref{fig:MV_f1}--\ref{fig:MV_f3} show the vacuum-energy/matter spectra
for the same three frequency bands.  Detecting stasis of this type is challenging due to the large suppression of the signal for all allowed values of $w_s$. This feature could be detectable for allowed values of $r$ in the fine tuned scenario in which $f_{\rm end}$ coincides with the minimum of
the BBO PLS curve. 

\section{Vacuum-energy/radiation stasis (\texorpdfstring{$-1/3 < \ws < 1/3$}{-1/3 < ws < 1/3})}
\label{sec:VR}

The vacuum-energy/radiation stasis scenario of~\cite{Dienes:2023ziv}
(Sec.~IV therein) covers $\ws\in(-1/3, 1/3)$, encompassing the canonical and
vacuum-energy/matter ranges.  The figures below therefore display the union of
the spectral shapes seen in the previous two subsections, smoothly interpolating
from the more suppressed ($\ws < 0$) to the less suppressed ($\ws > 0$) regimes.

\begin{figure}[t]
  \centering
  \includegraphics[width=1\linewidth]{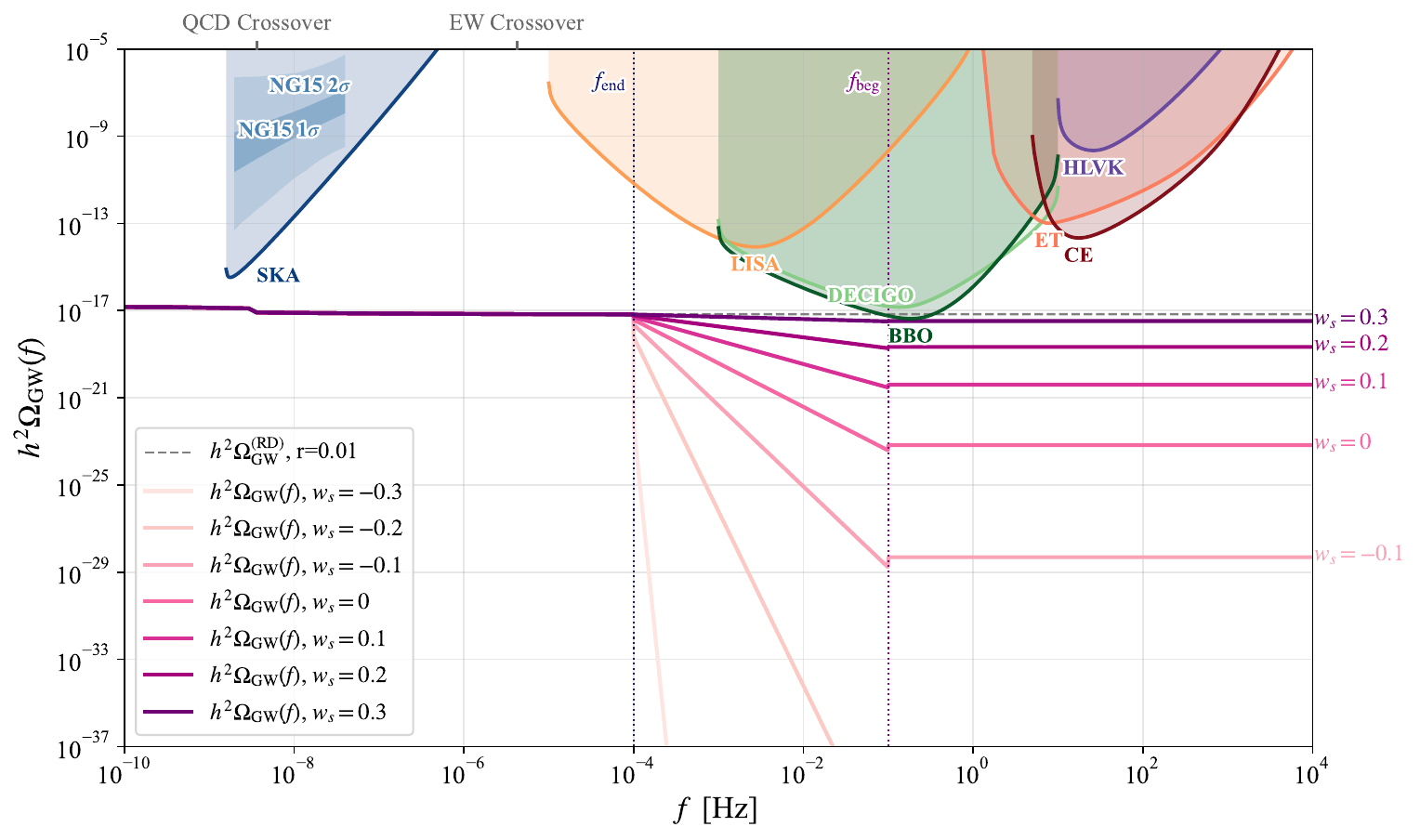}
  \caption{Stochastic GW background $h^2\Omega_{\rm GW}(f)$ for
    vacuum-energy/radiation stasis ($\ws\in(-1/3,1/3)$) in Band~I.
    Grey dashed: RD baseline at $r=0.01$.
    Shaded: PLS curves~\protect\cite{Schmitz:2020rag}.
    Blue band: NANOGrav 15-year posterior~\protect\cite{NANOGrav:2023gor}.}
  \label{fig:VR_f1}
\end{figure}

\begin{figure}[t]
  \centering
  \includegraphics[width=1\linewidth]{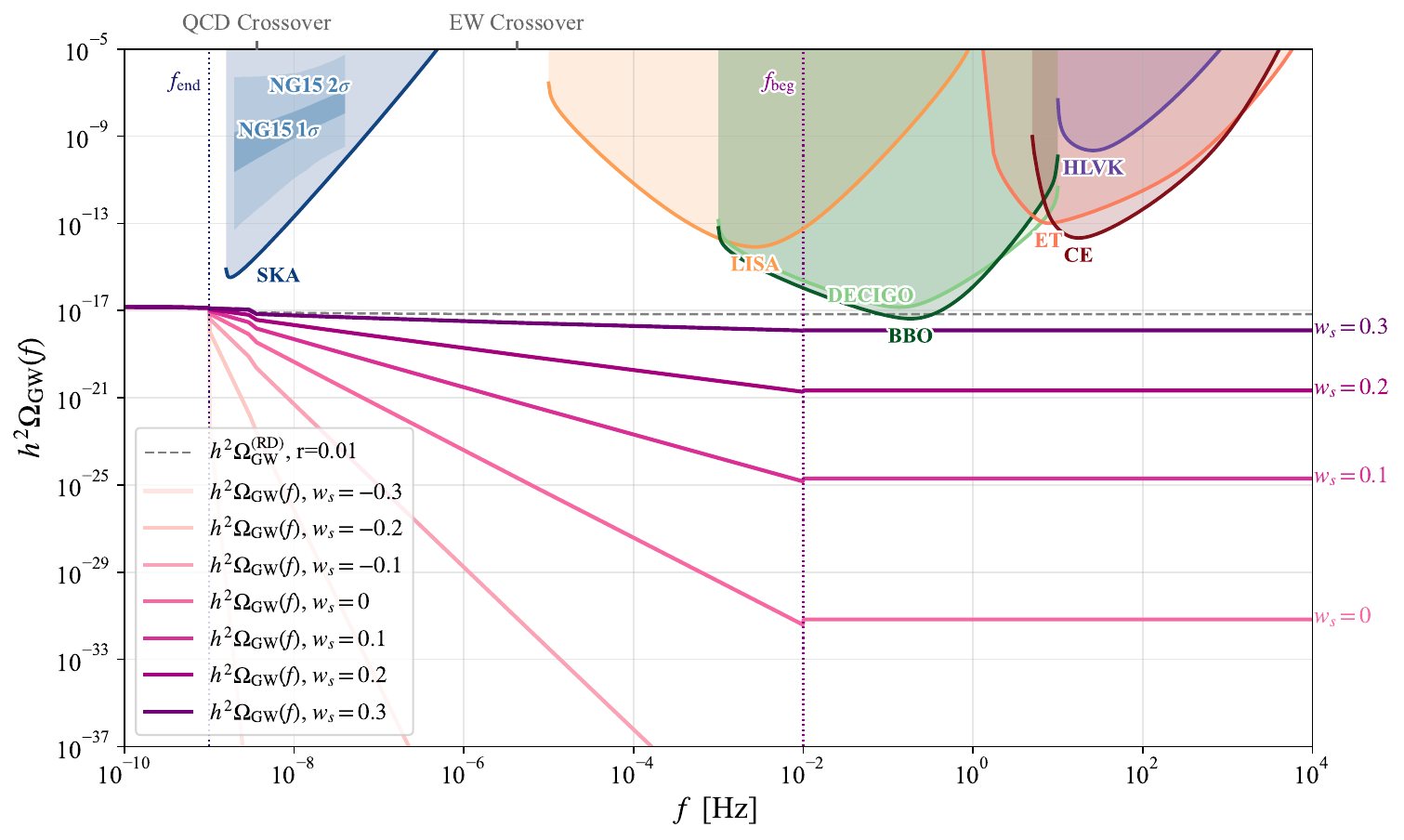}
  \caption{Same as Fig.~\ref{fig:VR_f1} for Band~II.
    The $g_*$ steps at $f_\mathrm{QCD}$ and $f_\mathrm{EW}$ are
    present within the stasis band for this configuration.}
  \label{fig:VR_f2}
\end{figure}

\begin{figure}[t]
  \centering
  \includegraphics[width=1\linewidth]{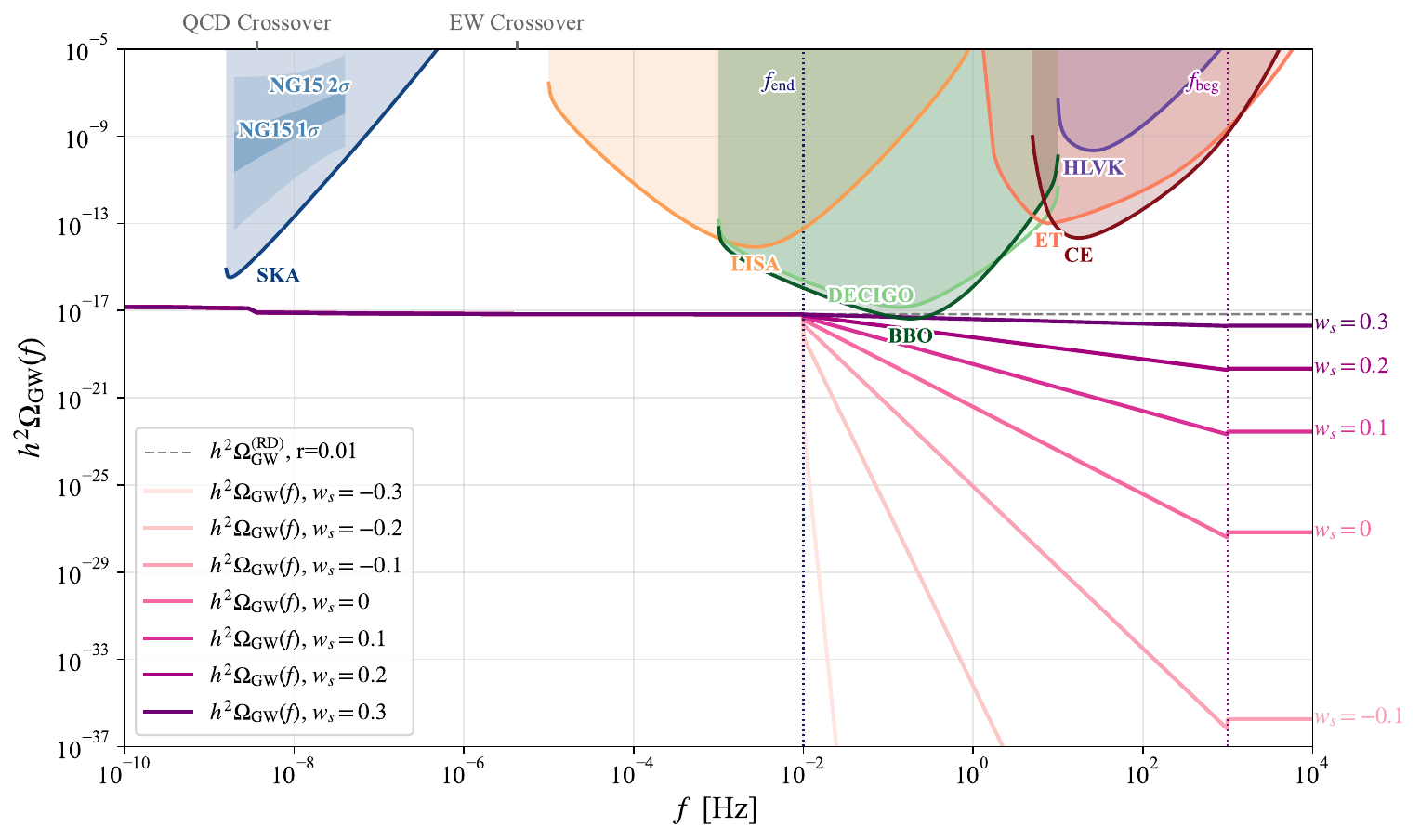}
  \caption{Same as Fig.~\ref{fig:VR_f1} for Band~III.
    The spectral template spans the full $\ws$ range,
    from strong suppression ($\ws\to -1/3$) to the flat RD baseline ($\ws = 1/3$).}
  \label{fig:VR_f3}
\end{figure}

Figures~\ref{fig:VR_f1}--\ref{fig:VR_f3} show the spectra.  The detectability
conclusions mirror those of the canonical and VE/M cases above: Band~III
placed in the BBO/DECIGO range is the most favorable configuration, with BBO
capable of detecting the stasis feature for $r\gtrsim 0.01$
for large values of $\ws$ within the allowed range.

\section{Additional plots: r = 0.036}

In this appendix we provide plots of $h^2\Omega_{\rm GW}(f)$ for all four models we cover in this paper for the tensor-to-scalar ratio r set to the upper limit provided by Planck, r = 0.036. 

\begin{figure}[t]
  \centering
  \begin{subfigure}{0.49\linewidth}
    \centering
    \includegraphics[width=\linewidth]{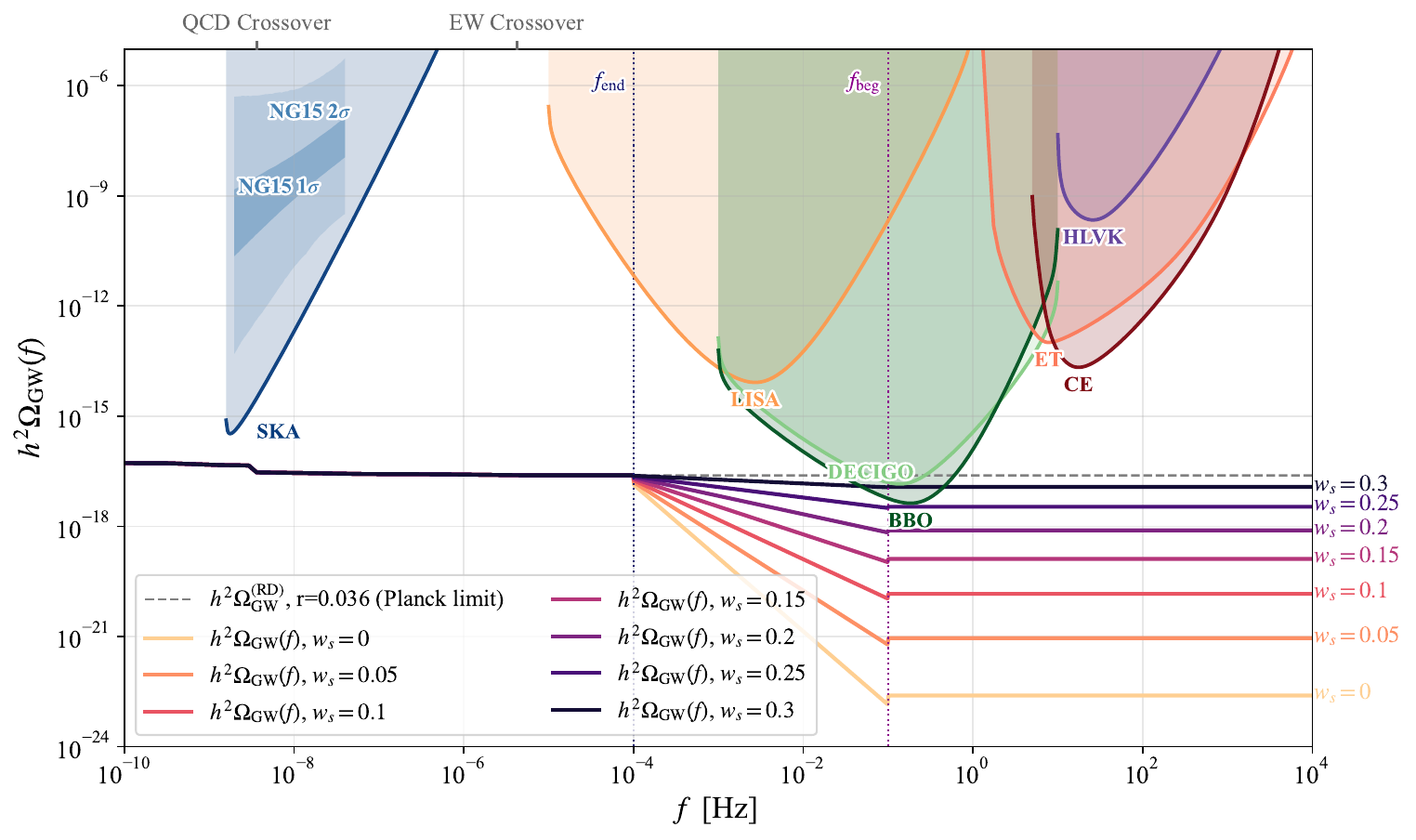}
    \caption{Band~I, $\fend=10^{-4}\,\mathrm{Hz}$, $\fbeg=10^{-1}\,\mathrm{Hz}$;
      $T_\mathrm{end}\approx 3.8\,\mathrm{TeV}$. }
    \label{fig:canon_f1_planck}
  \end{subfigure}
  \hfill
  \begin{subfigure}{0.49\linewidth}
    \centering
    \includegraphics[width=\linewidth]{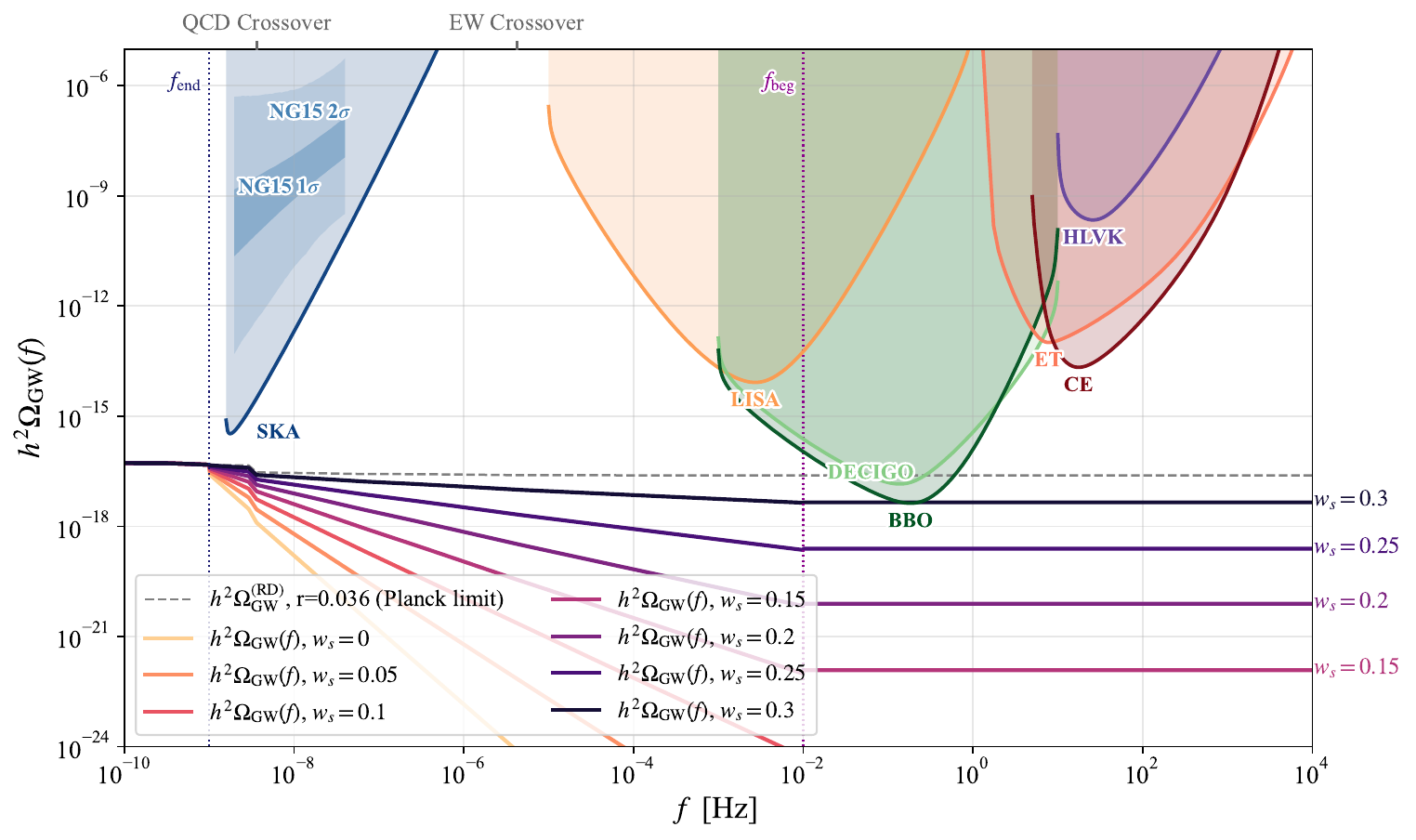}
    \caption{Band~II, $\fend=10^{-9}\,\mathrm{Hz}$, $\fbeg=10^{-2}\,\mathrm{Hz}$;
      $T_\mathrm{end}\approx 55\,\mathrm{MeV}$. }
    \label{fig:canon_f2_planck}
  \end{subfigure}

  \vspace{0.5em}

  \begin{subfigure}{0.49\linewidth}
    \centering
    \includegraphics[width=\linewidth]{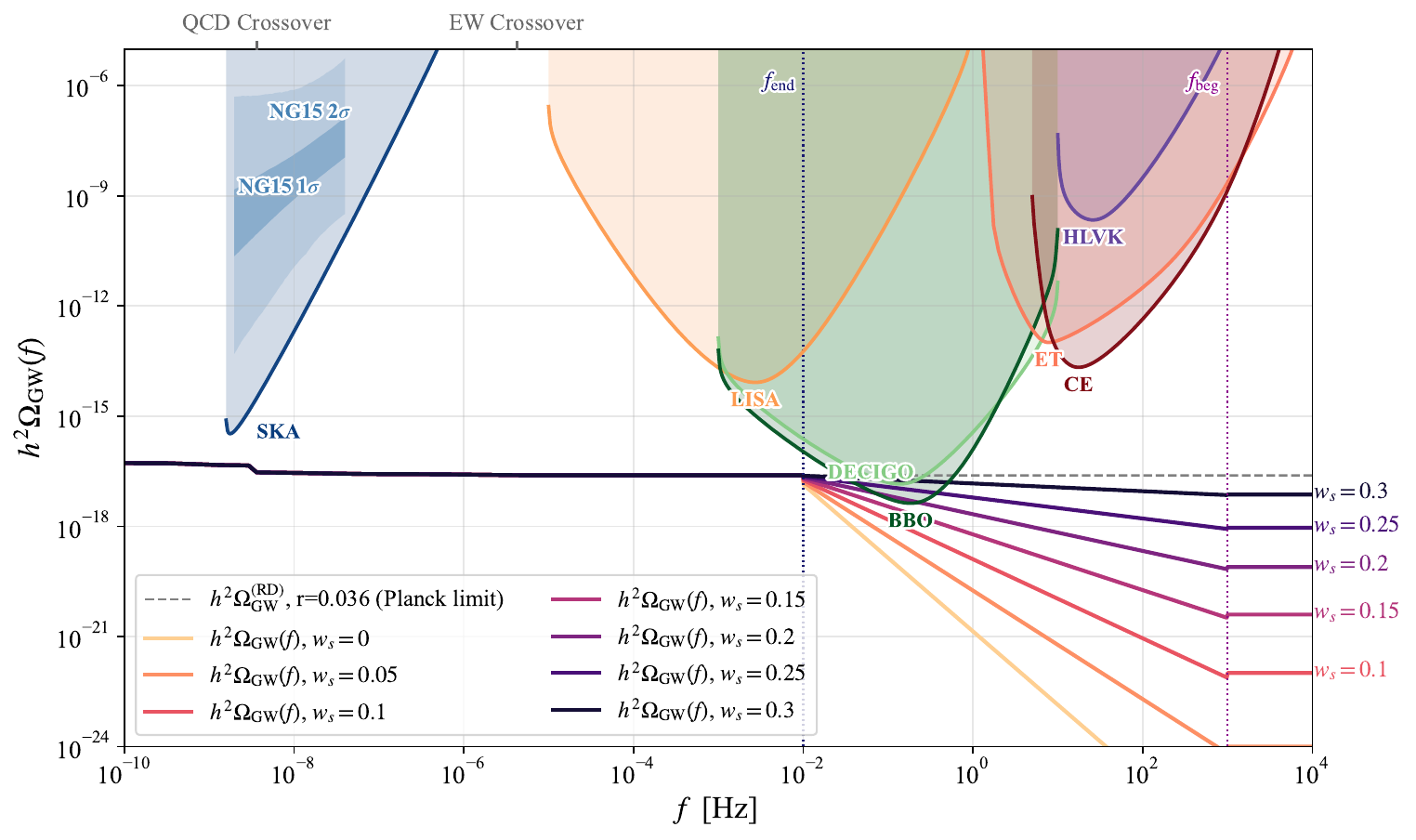}
    \caption{Band~III, $\fend=10^{-2}\,\mathrm{Hz}$, $\fbeg=10^{3}\,\mathrm{Hz}$;
      $T_\mathrm{end}\approx 380\,\mathrm{TeV}$.}
    \label{fig:canon_f3_planck}
  \end{subfigure}

  \caption{Stochastic GW background $h^2\Omega_{\rm GW}(f)$ for canonical
    stasis ($\ws\in[0,1/3]$) across three frequency bands.
    Grey dashed: RD at $r=0.036$ (Planck upper limit).
    Shaded: detector PLS curves~\protect\cite{Schmitz:2020rag} with
    $T_{obs}=4$ years and $\rho_{thr}=1$.
    Blue band: NANOGrav 15-year posterior~\protect\cite{NANOGrav:2023gor}.}
  \label{fig:canon_combined}
\end{figure}

\begin{figure}[t]
  \centering
  \begin{subfigure}{0.49\linewidth}
    \centering
    \includegraphics[width=\linewidth]{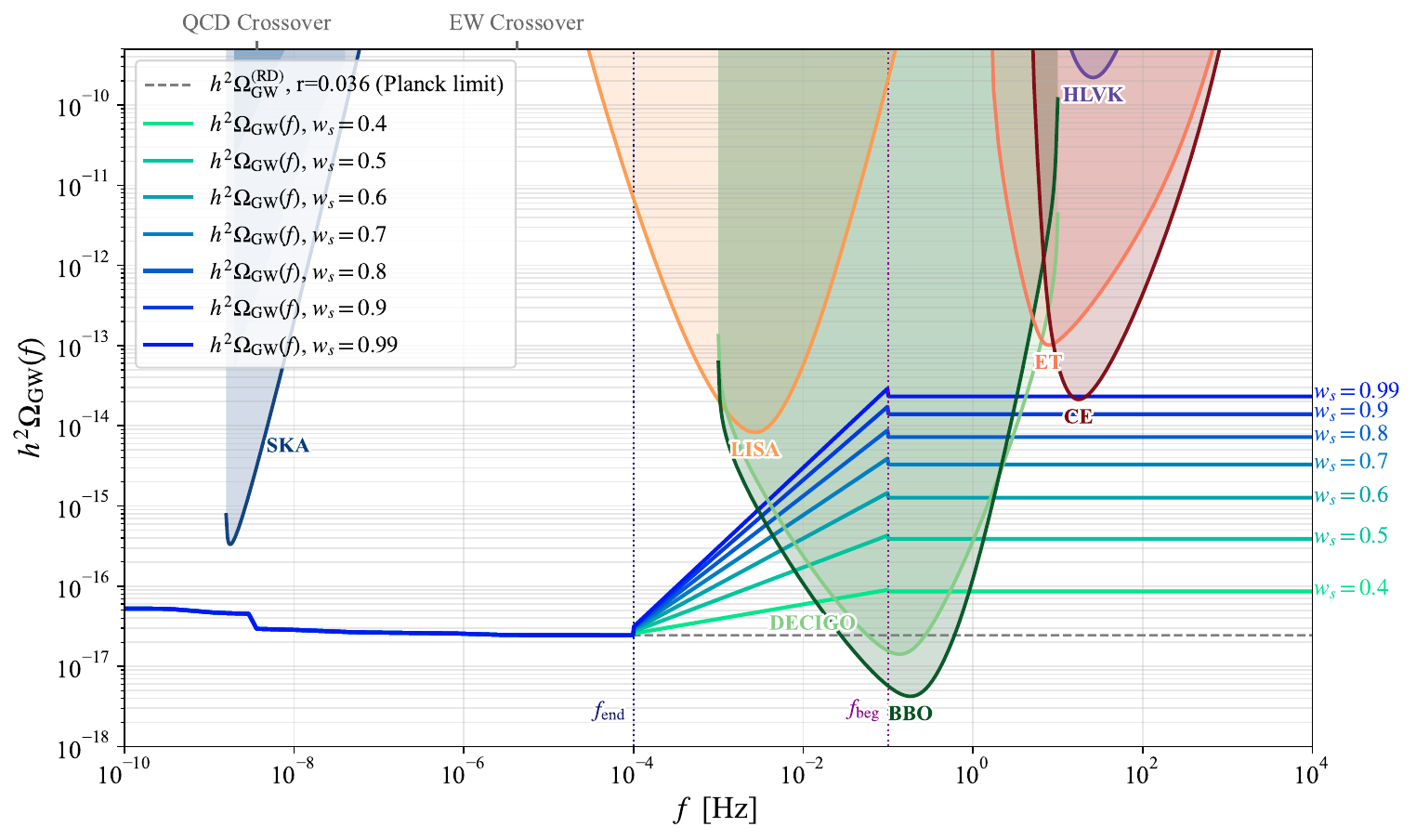}
    \caption{Band~I, $\fend=10^{-4}\,\mathrm{Hz}$, $\fbeg=10^{-1}\,\mathrm{Hz}$;
      $T_\mathrm{end}\approx 3.8\,\mathrm{TeV}$.}
  \end{subfigure}
  \hfill
  \begin{subfigure}{0.49\linewidth}
    \centering
    \includegraphics[width=\linewidth]{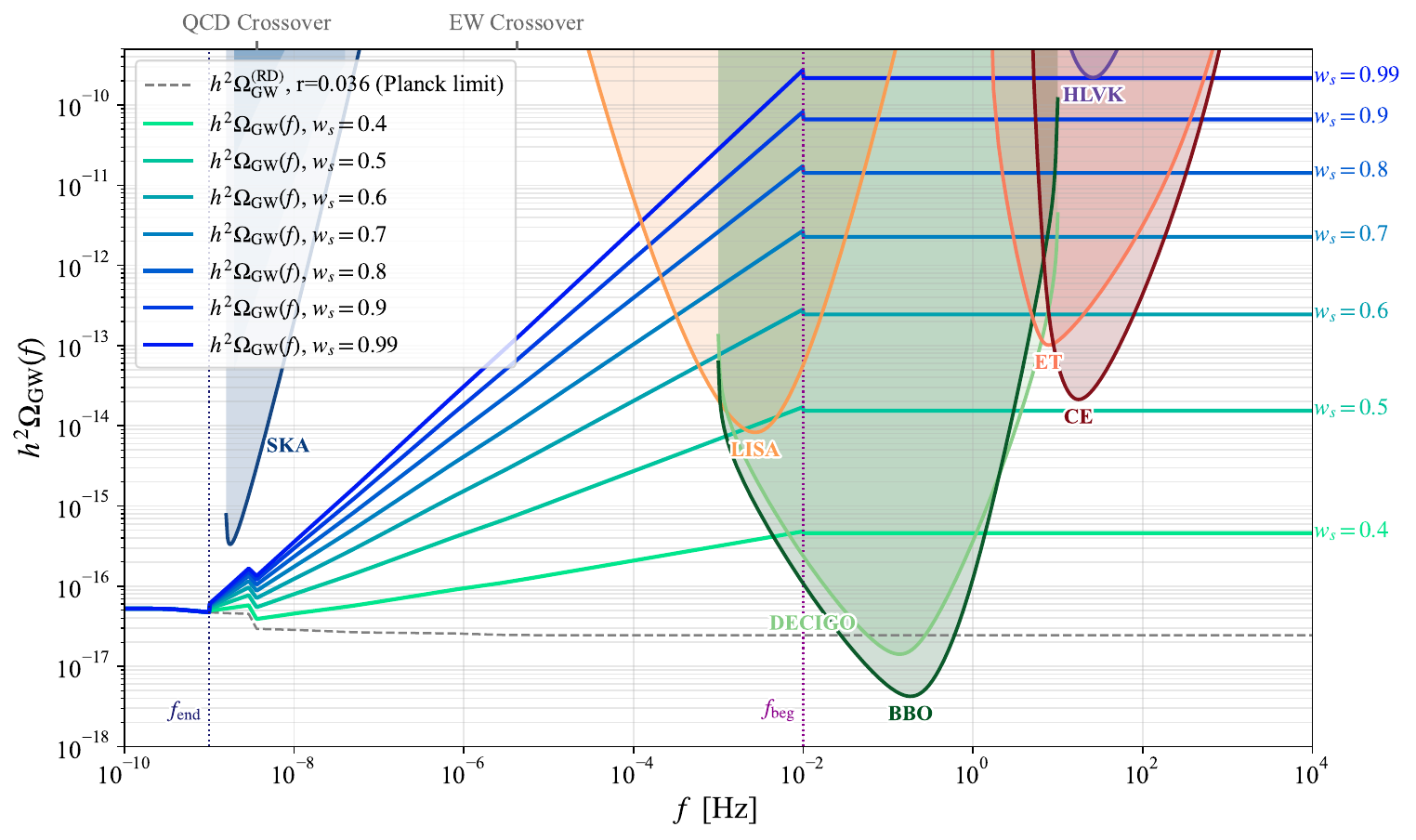}
    \caption{Band~II, $\fend=10^{-9}\,\mathrm{Hz}$, $\fbeg=10^{-2}\,\mathrm{Hz}$;
      $T_\mathrm{end}\approx 55\,\mathrm{MeV}$. }
  \end{subfigure}
  \vspace{0.5em}
  \begin{subfigure}{0.49\linewidth}
    \centering
    \includegraphics[width=\linewidth]{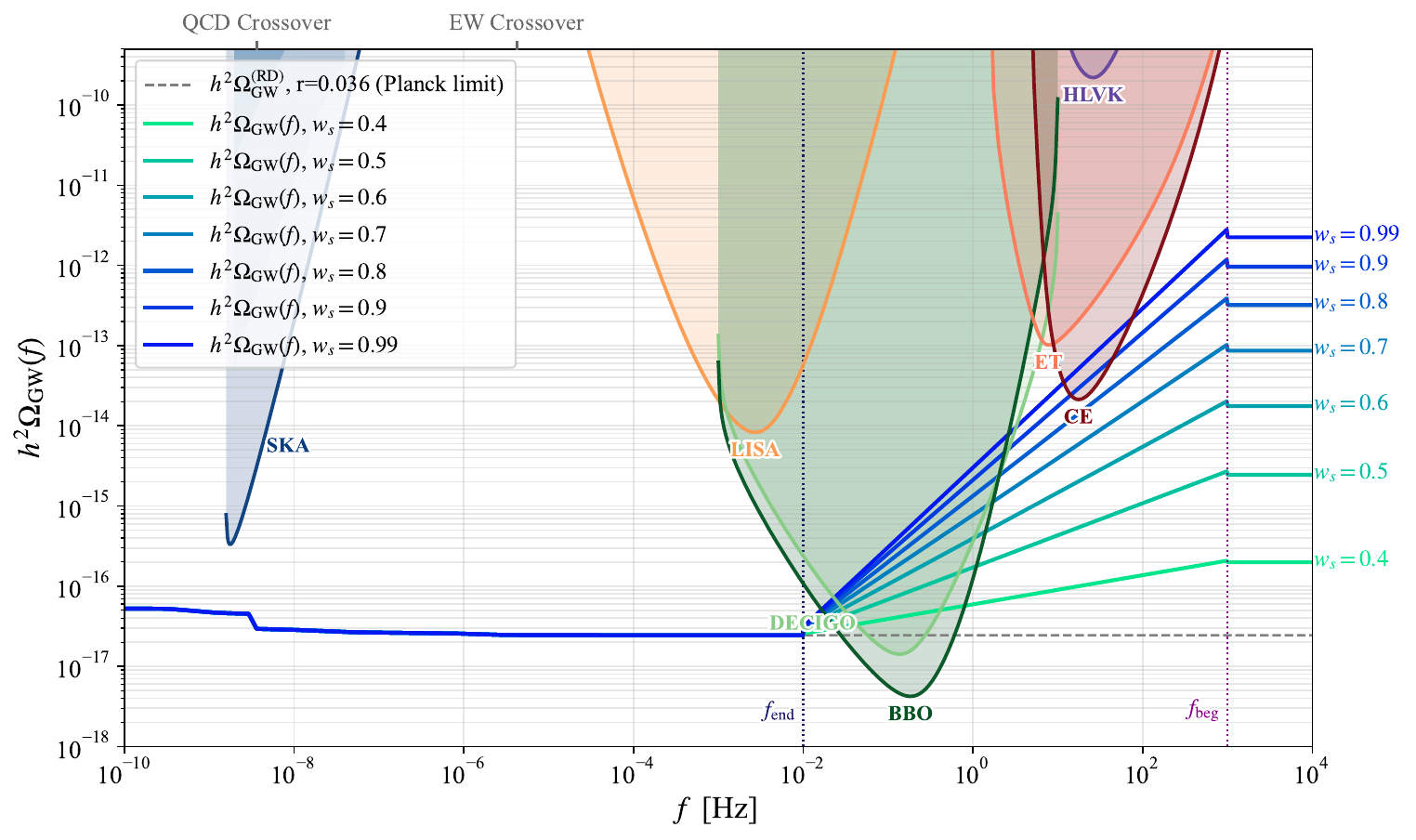}
    \caption{Band~III, $\fend=10^{-2}\,\mathrm{Hz}$, $\fbeg=10^{3}\,\mathrm{Hz}$;
      $T_\mathrm{end}\approx 380\,\mathrm{TeV}$.}
  \end{subfigure}
  \caption{Stochastic GW background $h^2\Omega_{\rm GW}(f)$ for dynamical
    scalar stasis ($\ws\in[1/3,1]$) across three frequency bands.
    Grey dashed: RD baseline at $r=0.036$ (Planck upper limit).
    Shaded: detector PLS curves~\protect\cite{Schmitz:2020rag} with
    $T_{obs}=4$ years and $\rho_{thr}=1$.
    Blue band: NANOGrav 15-year posterior~\protect\cite{NANOGrav:2023gor}.}
  \label{fig:DS_combined}
\end{figure}

\begin{figure}[t]
  \centering
  \begin{subfigure}{0.49\linewidth}
    \centering
    \includegraphics[width=\linewidth]{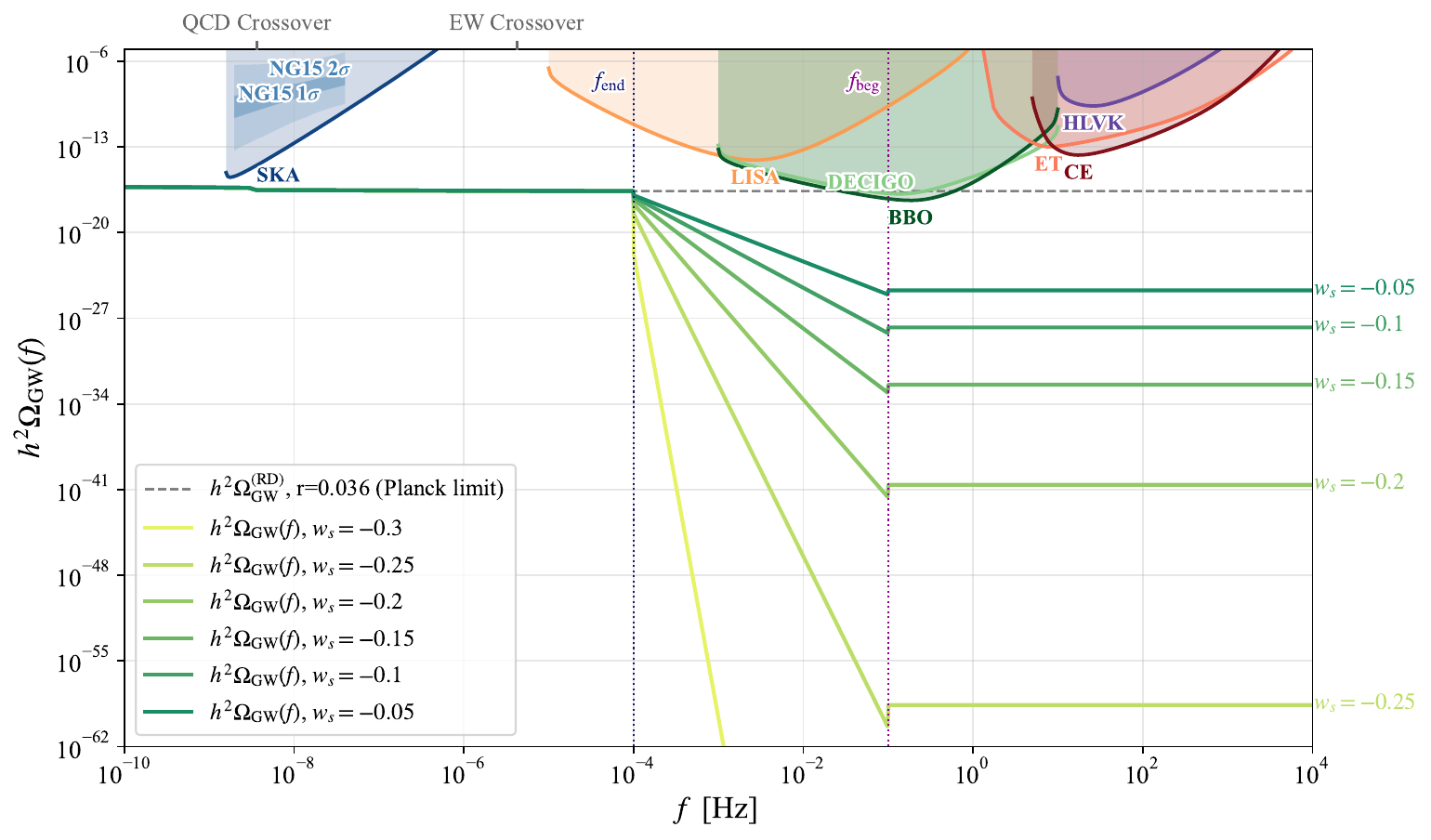}
    \caption{Band~I, $\fend=10^{-4}\,\mathrm{Hz}$, $\fbeg=10^{-1}\,\mathrm{Hz}$;
      $T_\mathrm{end}\approx 3.8\,\mathrm{TeV}$. }
  \end{subfigure}
  \hfill
  \begin{subfigure}{0.49\linewidth}
    \centering
    \includegraphics[width=\linewidth]{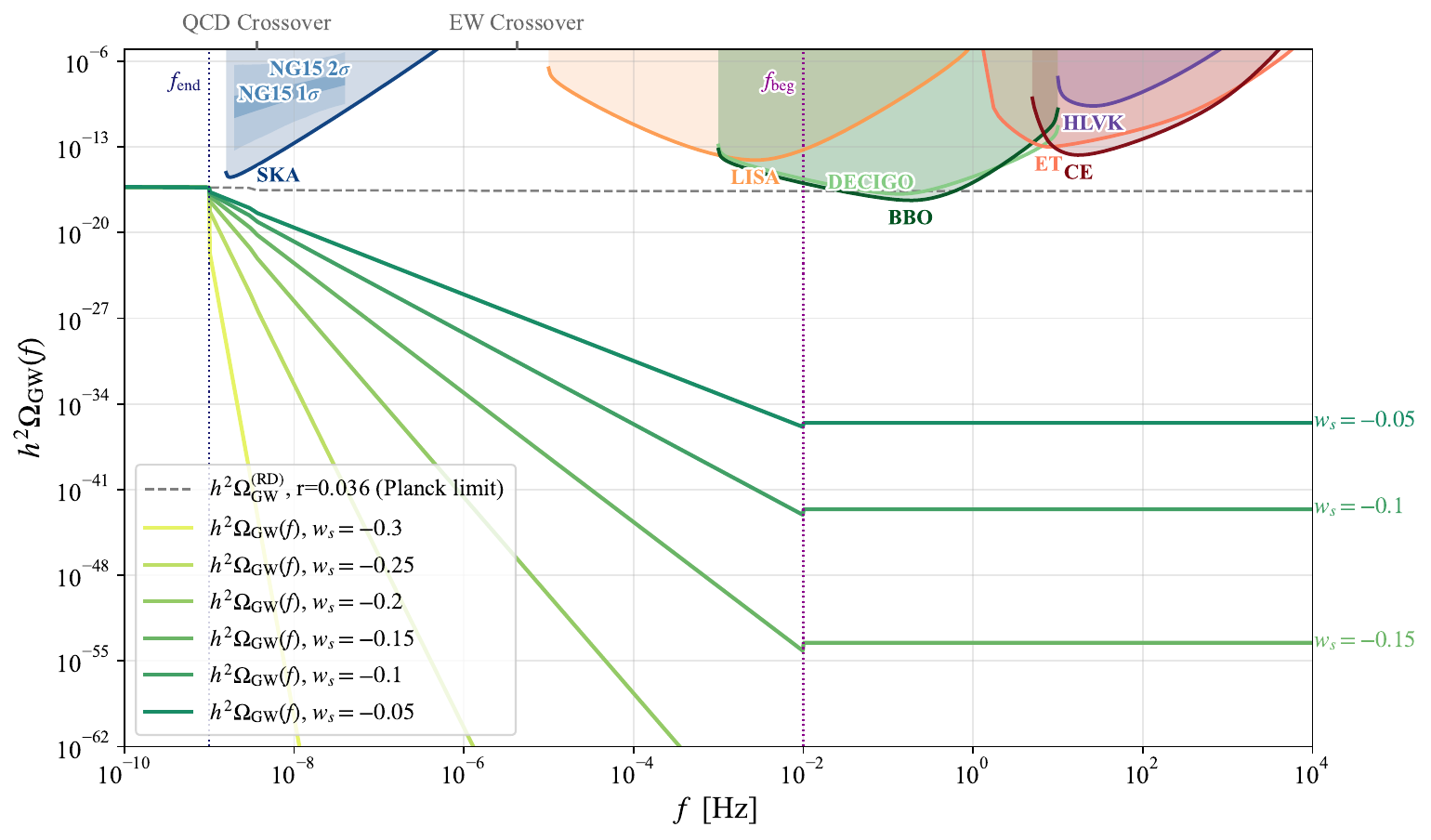}
    \caption{Band~II, $\fend=10^{-9}\,\mathrm{Hz}$, $\fbeg=10^{-2}\,\mathrm{Hz}$;
      $T_\mathrm{end}\approx 55\,\mathrm{MeV}$.}
  \end{subfigure}
  \vspace{0.5em}
  \begin{subfigure}{0.49\linewidth}
    \centering
    \includegraphics[width=\linewidth]{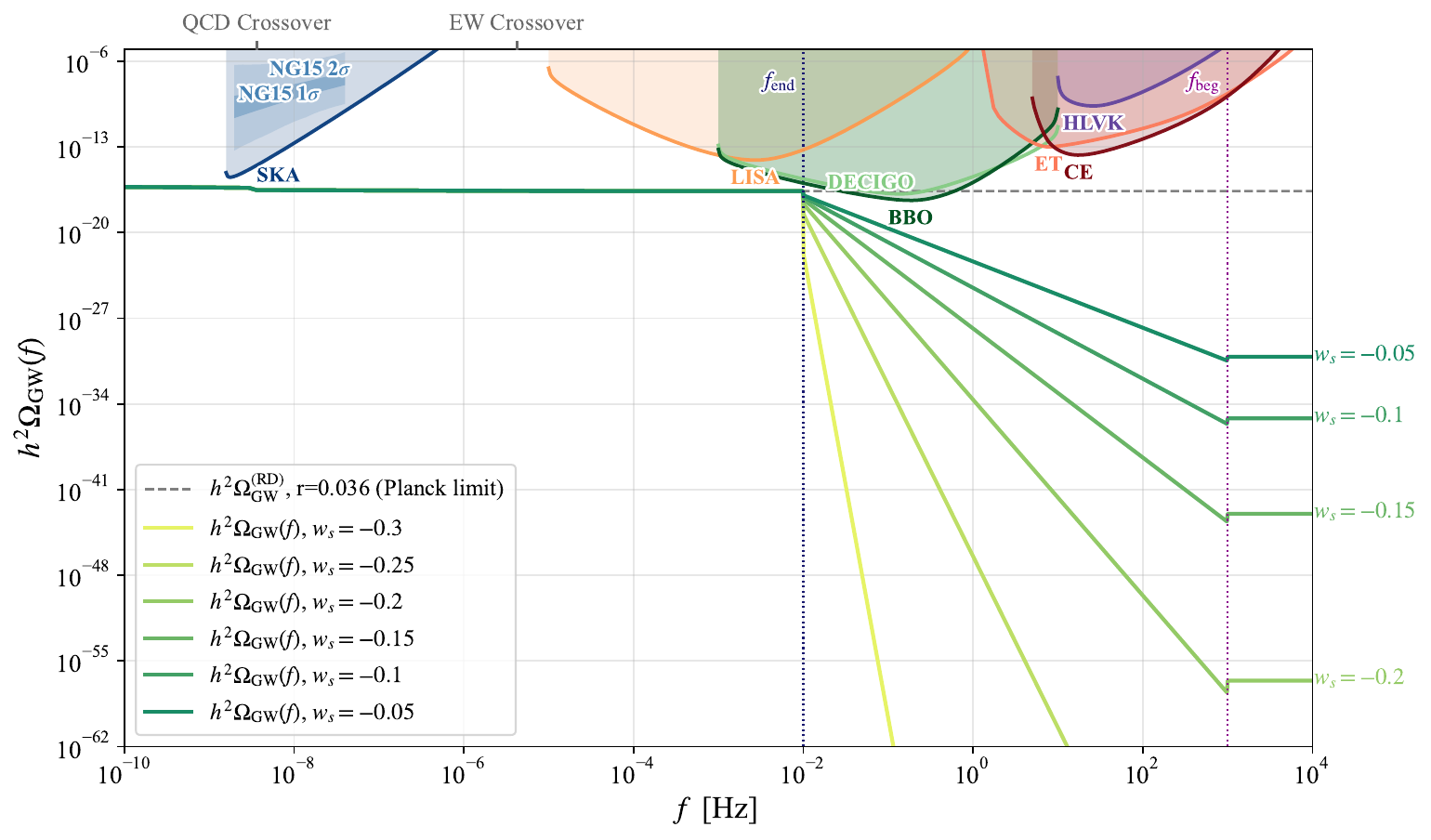}
    \caption{Band~III, $\fend=10^{-2}\,\mathrm{Hz}$, $\fbeg=10^{3}\,\mathrm{Hz}$;
      $T_\mathrm{end}\approx 380\,\mathrm{TeV}$.}
  \end{subfigure}
  \caption{Stochastic GW background $h^2\Omega_{\rm GW}(f)$ for
    vacuum-energy/matter stasis ($\ws\in(-1/3,0)$) across three
    frequency bands.
    Grey dashed: RD baseline at $r=0.036$ (Planck upper limit).
    Shaded: detector PLS curves~\protect\cite{Schmitz:2020rag} with
    $T_{obs}=4$ years and $\rho_{thr}=1$.
    Blue band: NANOGrav 15-year posterior~\protect\cite{NANOGrav:2023gor}.}
  \label{fig:MV_combined}
\end{figure}

\begin{figure}[t]
  \centering
  \begin{subfigure}{0.49\linewidth}
    \centering
    \includegraphics[width=\linewidth]{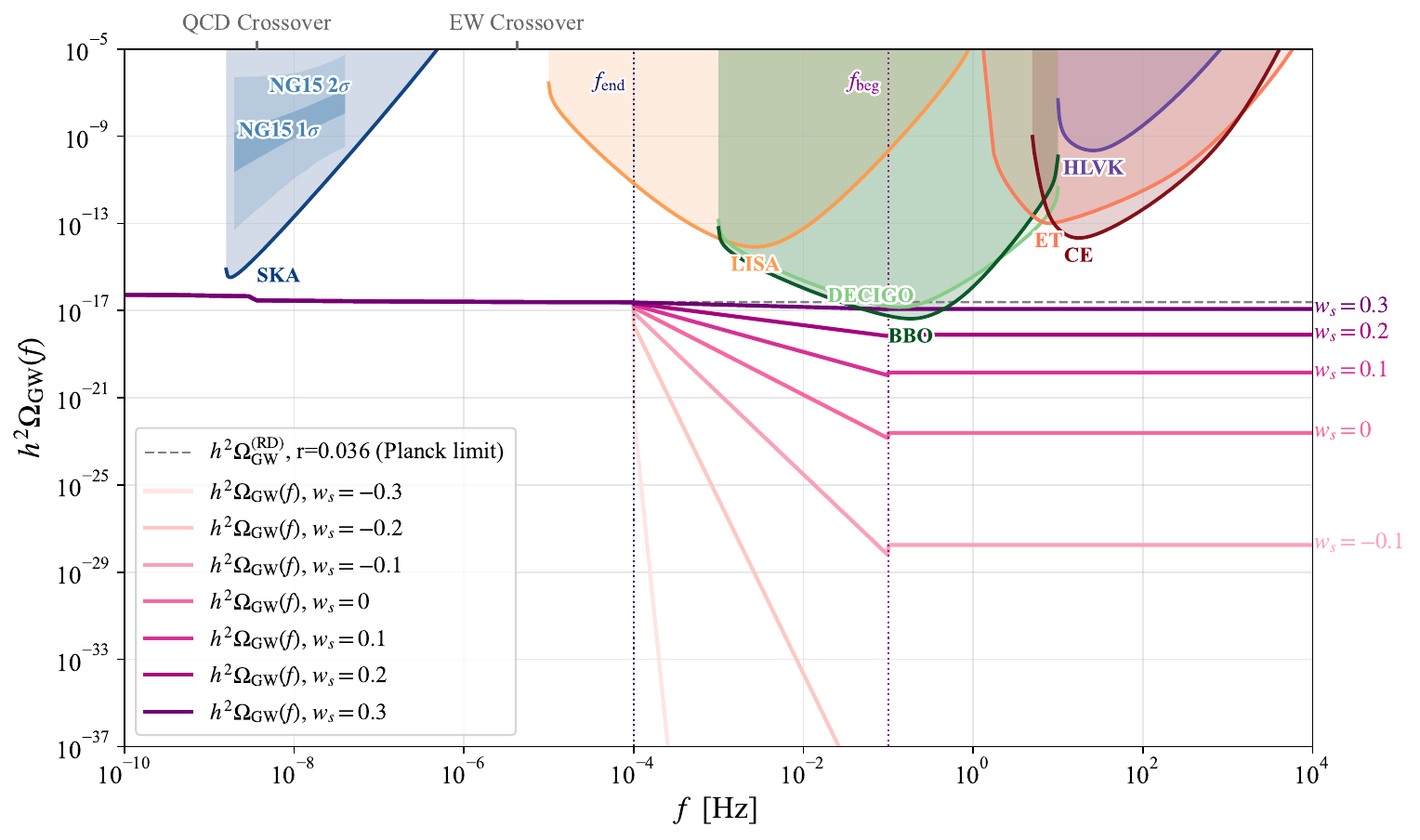}
    \caption{Band~I, $\fend=10^{-4}\,\mathrm{Hz}$, $\fbeg=10^{-1}\,\mathrm{Hz}$;
      $T_\mathrm{end}\approx 3.8\,\mathrm{TeV}$.}
  \end{subfigure}
  \hfill
  \begin{subfigure}{0.49\linewidth}
    \centering
    \includegraphics[width=\linewidth]{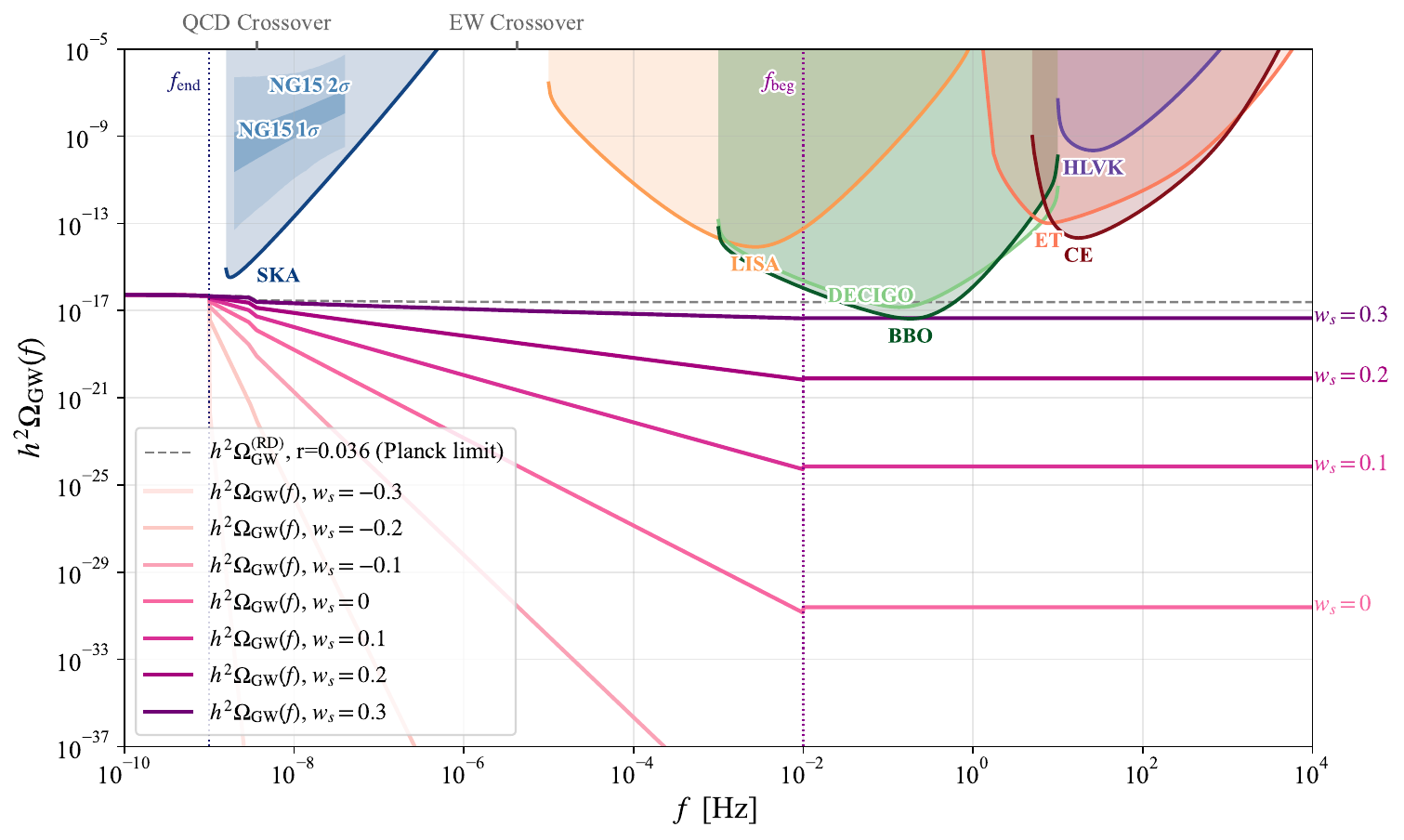}
    \caption{Band~II, $\fend=10^{-9}\,\mathrm{Hz}$, $\fbeg=10^{-2}\,\mathrm{Hz}$;
      $T_\mathrm{end}\approx 55\,\mathrm{MeV}$.}
  \end{subfigure}
  \vspace{0.5em}
  \begin{subfigure}{0.49\linewidth}
    \centering
    \includegraphics[width=\linewidth]{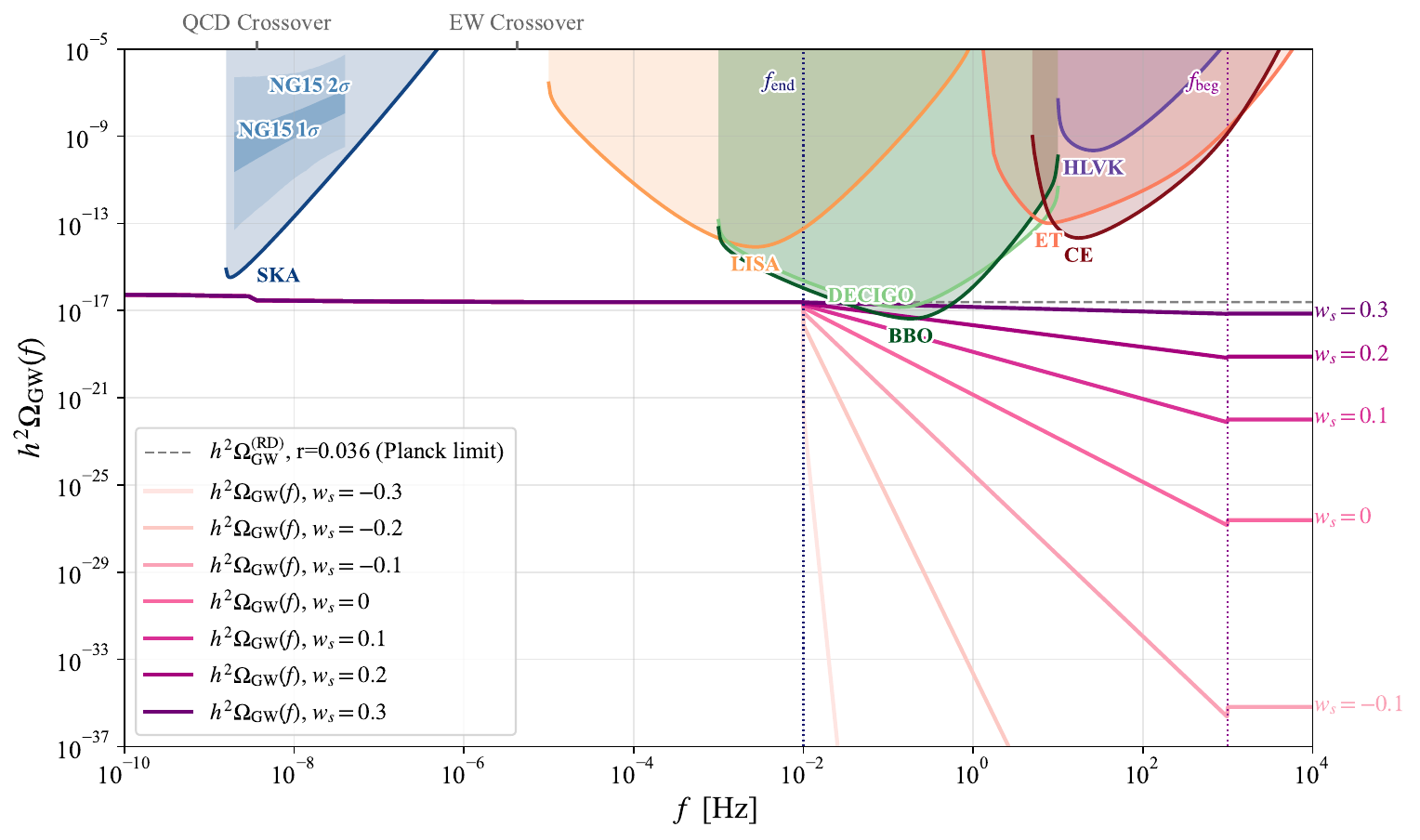}
    \caption{Band~III, $\fend=10^{-2}\,\mathrm{Hz}$, $\fbeg=10^{3}\,\mathrm{Hz}$;
      $T_\mathrm{end}\approx 380\,\mathrm{TeV}$.}
  \end{subfigure}
  \caption{Stochastic GW background $h^2\Omega_{\rm GW}(f)$ for
    vacuum-energy/radiation stasis ($\ws\in(-1/3,1/3)$) across three
    frequency bands.
    Grey dashed: RD baseline at $r=0.036$ (Planck upper limit).
    Shaded: detector PLS curves~\protect\cite{Schmitz:2020rag} with
    $T_{obs}=4$ years and $\rho_{thr}=1$.
    Blue band: NANOGrav 15-year posterior~\protect\cite{NANOGrav:2023gor}.}
  \label{fig:VR_combined}
\end{figure}

\end{document}